\documentclass[twocolumn,showpacs,preprintnumbers,amsmath,amssymb,eqsecnum]{revtex4}
\usepackage{graphicx}
\usepackage{subfigure}
\usepackage[dvips]{epsfig}
\usepackage{bm}
\usepackage{hyperref}
\def\cc{{\cal C}}

\usepackage{graphicx}
\usepackage{natbib}
\usepackage{amssymb}

\begin{document}

\title{Multipolar universal relations between $f$-mode frequency and tidal
deformability  of compact stars}

\author{T.~K. Chan\footnote{Present address: Department of Physics,
University of California at San Diego, 9500 Gilman Drive, La
Jolla, CA 92093, USA. Email address: chantsangkeung@gmail.com},
Y.-H. Sham\footnote{Email address: yhsham@phy.cuhk.edu.hk},
P.~T. Leung\footnote{Email address: ptleung@phy.cuhk.edu.hk}, and
L.-M. Lin\footnote{Email address: lmlin@phy.cuhk.edu.hk} }

\affiliation{Department of Physics and Institute of Theoretical
Physics, The Chinese University of Hong Kong, Shatin, Hong Kong
SAR, China }

\date{\today}

\begin{abstract}
Though individual stellar parameters of compact stars usually
demonstrate obvious dependence on the equation of state (EOS),
EOS-insensitive universal formulas relating these parameters
remarkably exist. In the present paper, we explore the
interrelationship between two  such formulas, namely the $f$-$I$
relation connecting the $f$-mode quadrupole  oscillation frequency
$\omega_2$ and the moment of inertia $I$, and the $I$-Love-$Q$
relations relating $I$, the quadrupole tidal deformability
$\lambda_2$, and the quadrupole moment $Q$, which have been
proposed by Lau, Leung, and Lin [Astrophys. J. {\bf 714}, 1234 (2010)] and
Yagi and Yunes [Science {\bf 341}, 365 (2013)], respectively. A
relativistic  universal relation between $\omega_l$ and
$\lambda_l$ with the same angular momentum $l=2,3,\ldots$, the
so-called ``diagonal $f$-Love relation" that holds for realistic
compact stars and stiff polytropic stars, is unveiled here. An
in-depth investigation in the Newtonian limit is further carried
out to pinpoint its underlying physical mechanism and hence leads
to a unified $f$-$I$-Love relation. We reach the conclusion that these
EOS-insensitive formulas stem from a common physical origin ---
compact stars can be considered as quasiincompressible when they
react to slow time variations introduced by $f$-mode oscillations,
tidal forces and rotations.
\end{abstract}



\maketitle

\section{Introduction}
\label{sec:intro} The structure of neutron stars (NSs), the
remnants of supernovae, is dictated by the strong gravity
prevailing inside such stars. As a result, the density of the
inner core of a typical NS can be several times the normal nuclear
density $ 2.8 \times 10^{14}\, {\rm g \,cm}^{-3}$, which is not
yet achievable (at least in a stable form) in the terrestrial
environment. The equation of state (EOS) of nuclear matter in NSs
is often masked by various uncertainties in nuclear and particle
physics, leading to the associated uncertainties in the physical
characteristics of NSs, including their masses, radii, and moments
of inertia. Thus, it has become a common practice for nuclear
physicists to examine  how EOSs of high-density nuclear matter
could affect the structure of NSs, e.g., the mass-radius relation,
and systematic investigations along such direction have also been
carried out (see, e.g.,
\citep{Weber-pulsar,Lattimer:2001,Lattimer:2005p7082,lattimer2007nso}).
On the other hand, mass efforts have been pooled together to infer
the details of nuclear matter from various possibly detectable
characteristics (such as mass, radius, moment of inertia and
gravitational-wave spectrum) of NSs (see, e.g.,
\citep{Andersson1996,Andersson1998,Lattimer:2001,Lattimer:2005p7082,lattimer2007nso,Ozel_09_PRD,2013arXiv1303.3282H,Latt_14_EPJ,Lattimer_2013_ApJ}).

However, paralleling these attempts to differentiate the structure
of NSs with respect to nuclear EOS and vice versa,  several
universal EOS-insensitive formulas connecting different physical
parameters (e.g., mass, radius and moment of inertia) of NSs or
quark stars (QSs) have also been discovered. Such formulas are
important and useful as they enable astronomers to infer (or at
least to constrain) a physical parameter of NSs (or QSs) from the
others that are more amenable to physical measurement even in the
absence of exact information of relevant EOSs. For example,
Lattimer and Schutz~\cite{Lattimer:2005p7082} and 
Bejger and Haensel~\cite{Bejger:2002p8392} found
formulas relating the moment of inertia $I$, the mass $M$, and the
radius $R$ for NSs constructed with different realistic EOSs,
which could be used to determine the moment of inertia of star A
in the double pulsar system J0737-3039
\citep{Burgay:2003p531,Lyne:2004p1153} to about 10\% accuracy.
Besides, several universal behaviors of the quadrupolar $f$-mode
oscillations of NSs have been established and applied to infer the
EOS of nuclear matter
\citep{Andersson1996,Andersson1998,Benhar1999:p797,
Benhar:2004:p124015,Tsui05:1,Tsui05:2,Tsui05:3,Tsui:prd,Lau:2010p1234},
and similar studies have been extended to the quadrupolar and
higher spherical order $f$-mode  oscillations of rapidly rotating
neutron stars \citep{Kokk_rotating,Kokk_rotating_real}

It is particularly interesting to note that  some of these
universal formulas actually relate the dynamical response of a NS
(or QS) under external perturbation to its static stellar
structure. In particular, Lau, Leung, and Lin~\cite{Lau:2010p1234} 
found that the frequency and damping rate of $f$-mode quadrupole oscillation 
are expressible in terms of $M$ and the effective compactness $\eta
\equiv \sqrt{M^3/I}$, hereafter referred to as the ``$f$-$I$
relations". Hence, the values of $M$, $R$, and $I$ of a NS (or QS)
can be inferred accurately from the $f$-mode gravitational-wave
signals \citep{Lau:2010p1234}. Since pulsating NSs are expected to
be promising sources of gravitational waves, the above-mentioned
relations can lead to useful information about the static
structure of NSs (or QSs) once the third-generation gravitational-wave 
detectors (e.g., the Einstein Telescope
\citep{Einstein_tele}) are available in the future.

On the other hand,  universal relations expressing the distortion
of a NS induced by tidal forces or spin in terms of its static
structural parameters have also been found recently
\citep{Yagi:2013long,Yagi:2013,Urbanec:13p1903,Baubock_13_ApJ}.
The so-called ``$I$-Love-$Q$ relations'', discovered by
Yagi and Yunes~\cite{Yagi:2013long,Yagi:2013}, relate $I$, 
the quadrupole tidal
Love number $\lambda_{2 }$ (or, more precisely, tidal
deformability \citep{Damour:09p084035,Yagi_14_PRD}), and the
spin-induced quadrupole moment  $Q$, with $M$ being a proper
scaling parameter. The relations are robust and also prevail
in several different  situations,
including binary systems with strong dynamical tidal field
\citep{Maselli:2013}, magnetized NSs with sufficiently high
rotation rates \citep{Haskell_14_mnras}, and rapidly rotating
stars \citep{Pappas_14_prl,Chak_14_prl}. Besides, there are works
extending the $I$-Love-$Q$ relations to consider the relation between
higher-order multipole moments induced by either tidal forces or
rotation \citep{Yagi_14_PRD,Yagi_hair_GR,Stein_hair_apj}.

Similar to other universal relations, these $I$-Love-$Q$ relations
enable us to infer any two of $I$, $\lambda_{2}$ and $Q$ from the
detected value of the remaining one even in the absence of prior
knowledge of the EOS of NSs. In addition, the I-Love-Q relations
could facilitate the analysis of gravitational-wave signals
emitted during the late stages of NS-NS binary mergers
\citep{Flanagan:08p021502,Hinderer:08p1216,Yagi:2013long,Yagi:2013},
and also serve as an indicator to identify the validity of other
modified gravity theories
\citep{Yagi:2013long,Yagi:2013,Sham_14_ApJ,Pani_Berti2014,Doneva_14_apj,Kleihaus_PRD_14}.

Despite the fact that the above-mentioned $f$-$I$ relation and
$I$-Love-$Q$ relations were discovered separately, the moment of
inertia $I$ is involved in both of them. It is physically natural
to expect that there should be a link between these two relations.
The present paper aims at examining their interrelationship and
establishing universal relations that can directly link the
characteristics of multipolar distortions induced respectively by
oscillations and tidal forces together. First of all, we
investigate the so-called ``$f$-Love relation'' between the $l$th
multipolar {\it f}-mode oscillation frequency $\omega_l$ and the
$l'$th multipolar tidal deformability $\lambda_{l'}$, where
$l,l'=2,3,4,\ldots$. We show that there exist approximately
EOS-independent and relativistically correct relations between
these two if $l=l'$, which are referred to as the diagonal $f$-Love
relations hereafter. It is obvious that the case with $l=2$ is a
direct consequence of the original $f$-$I$ and $I$-Love-$Q$ relations
\citep{Yagi:2013long,Yagi:2013,Lau:2010p1234}. For the
off-diagonal $f$-Love relations with $l \ne l'$, the relation
between the two relevant physical quantities become more
EOS-sensitive (see Sec.~III and Figs.~1-4). This finding is
important and useful. On the one hand, assuming that the mass of a
NS is known from astronomical measurements, one can use the
diagonal $f$-Love relation to determine the {\it f}-mode frequency
from the measured tidal deformability in the same angular momentum
sector
\citep{Flanagan:08p021502,Hinderer:08p1216,Yagi:2013long,Yagi:2013}
(and vice versa) even in the absence of detailed knowledge of
nuclear EOS. On the other, the details of EOS could be inferred
from the off-diagonal $f$-Love relations.

To pursue the physical origin of the EOS-independency of the
diagonal $f$-Love relations, we compare the $f$-Love relations of
realistic stars and polytropic stars with those of incompressible
stars (see Sec.~III and Figs.~1-4). We find that the latter can
nicely approximate the behavior of realistic stars (including both
NSs and QSs) and polytropic stars whose polytropic indices $N$ are
less than 1. As a matter of fact, the polytropic index of most (if
not all) popular realistic nuclear EOSs in the high density regime
is less than 1. Hence, these prevailing  EOSs are stiff enough to
guarantee the $f$-Love relations  except for NSs with very low
masses. If softer nuclear EOSs were proposed and used to construct
NSs, such stars would demonstrate significant deviations from  the
$f$-Love relation discovered in the present paper. Conversely, if
the $f$-Love relations were violated, it would hint at the softening
of nuclear matter due to physical mechanisms such as kaon
condensation or other phase transitions.

Furthermore, regarding the physical origin of the multipolar $f$-$I$,
$I$-Love and $f$-Love formulas in the Newtonian limit, we consider a
model star characterized by a density profile
$\rho(r)=\rho_0(1-\delta x^2)$ with $x \equiv r/R$ and $\rho_0$
being the central density (see Sec.~IV~A). Here $1 \geq \delta
\geq 0$ is a parameter controlling  the stiffness of the star.
With such a stellar model, we can approximately reproduce the
behavior of QSs and NSs whose polytropic index is less than 1,
while rendering calculations of physical quantities more
manageable and simplified (see the discussion in Sec.~IV~A).
First of all, we analytically derive the $f$-$I$ formula relating
$\omega_l$ and $I_n$, where $I_n \equiv \int_0^R \rho(r) r^{2+n}
\mathrm{d}r $ is the $n$th mass moment (see Sec.~IV~B), and
the $I$-Love formula connecting $I_n$ and $\lambda_l$ (see
Sec.~IV~C). In general, these two relations explicitly depend
on the value of $\delta$, i.e., the underlying EOS of the star,
and lack of universality. However, if $n=2l-2$, to first order in
$\delta$ both of them are independent of $\delta$. Consequently,
combining the $f$-$I$ and $I$-Love formulas, we can show that the
diagonal $f$-Love formula is also to first order independent of
$\delta$, corroborating the EOS independence of the diagonal
$f$-Love formula discovered here (see Sec.~IV~D).

The analytic studies developed in the present paper clearly
explain why these universal formulas are so insensitive to changes in EOSs.
We arrive at the conclusion  that the crux of the observed
universality is (i) these universal relations actually derive
from the behavior of incompressible stars;  (ii) they are
stationary with respect to changes in $N$ around the
incompressible limit (i.e. $N=0$) and hence are EOS-insensitive;
and (iii) the prevailing EOSs for nuclear matter are stiff enough
to be considered as almost incompressible (termed as
quasiincompressible here) in certain slow physical processes
(e.g., $f$-mode oscillations, tidal and rotational deformation).

Physically speaking, if an external perturbation is applied to
a compact star, the response of the star to such perturbation is
given by the Green's function, which consists of the contributions
arising from different oscillation modes
\citep{Chan1963,Press_77_ApJ}. Tidal deformation of NSs is merely
the zero-frequency component of the Green's function. We show in
the present paper that for quasiincompressible stars in the
Newtonian limit, the tidal field couples only to the {\it f}-mode
oscillation and establish a robust $f$-$I$-Love relation. Such a
relation readily shows that the $f$-$I$ and $I$-Love relations imply
each other, thus unifying these two independently discovered
universal relations.

The plan of the paper is as follows. In Sec.~\ref{review}
 we briefly review the  previously discovered universal behaviors of
the quadrupole $f$-$I$ and $I$-Love-$Q$ relations for NSs. We present the
multipolar $f$-Love universal relations in Sec.~\ref{f-Love}
and study how the stiffness of nuclear matter could affect the
accuracy of the $f$-Love relation. Newtonian analytic analysis will
be carried out in Sec.~\ref{theory} to establish the multipolar
$f$-$I$, $I$-Love and $f$-Love relations. A unified Green's function
approach to multipolar tidal deformation in the Newtonian regime
and hence a novel multipolar $f$-$I$-Love relation are given in
Sec.~\ref{f-I-Love}. The conclusions of the present paper are
presented in Sec.~\ref{sec:conclude}. Unless otherwise stated
explicitly, we use geometric units where $G=c=1$.

\section{Quadrupole $F$-$I$ and $I$-Love-$Q$ relations}
\label{review} 
When compact stars are perturbed away from their
equilibrium state, their subsequent oscillations can be analyzed
in terms of quasinormal modes (QNMs)
\citep{Press_1971,Leaver_1986,rmp,Kokkotas_rev,Berti:2009kk}. Each
QNM is characterized by  a complex eigenfrequency $\omega =
\omega_{\rm r} + i \omega_{\rm i}$, with $\omega_{\rm i}$
measuring its decay rate due to emission of gravitational waves.
For typical NSs, the oscillation frequency $\omega_{\rm r}$ of the
fundamental ({\it f}) mode usually lies in the kilohertz range,
which is lower than other QNMs such as the pressure ($p$) modes
and the spacetime ({\it w}) modes. Hence, as far as gravitational
waves emitted from oscillating NSs are concerned, {\it f}-mode
oscillations are most likely to be detectable with advanced
gravitational-wave observatories such as the Einstein telescope
\citep{Einstein_tele}. Owing to this, numerous approximate
formulas attempting to relate $\omega_{\rm r}$ and $\omega_{\rm
i}$ of $f$-mode oscillation to other physical parameters of NSs
have been proposed \citep{Andersson1996,Andersson1998,
Benhar1999:p797,Benhar:2004:p124015,Tsui05:1}. In most of these
cases, $M$ and $R$ were used as the independent parameters to
express the $f$-mode frequency. It is not until Lau, Leung, and 
Lin~\cite{Lau:2010p1234} introduced $M$ and $I$ as the two parameters
to quantify quadrupolar $f$-mode oscillations and found two nearly
EOS-independent relations
 \begin{eqnarray}
\bar{\omega}_2 \equiv M \omega_{\rm r}   &=& - 0.0047 + 0.133 \eta
+ 0.575 \eta^2 , \label{eq:fmode_real}\\ I^2 \omega_{\rm i} /M^5
&=& 0.00694 - 0.0256 \eta^2 , \label{eq:fmode_img}
\end{eqnarray}
where the effective compactness $\eta \equiv \sqrt{M^3 / I }$
replaces the role of the traditionally defined compactness ${\cal
C}\equiv M/R $. [In (\ref{eq:fmode_real}) we have corrected a
typographical error in \citep{Lau:2010p1234}.] As shown in Table 1
of \citep{Lau:2010p1234}, Eqs.~(\ref{eq:fmode_real}) and
(\ref{eq:fmode_img}) are more accurate than other universal
relations using $R$ as a parameter. The typical deviation  from
 the above two formulas for a realistic NS  is less than a few percent. 
In the present discussion we focus our attention on (\ref{eq:fmode_real}) and
refer it as the $f$-$I$ relation.

For a star with a given mass $M$, the effective radius $R_{\rm e}
\equiv \sqrt{I/M}$ measures its average size weighted by its mass
distribution and is therefore more relevant to the dynamics of the
star than the geometric radius $R$. Hence, the introduction of the
effective compactness $\eta = M/R_{\rm e}$ is expected to be a
crucial reason to lead to the better performance of
(\ref{eq:fmode_real}) and (\ref{eq:fmode_img}).

On the other hand, owing to the frequency sensitivity limit of
currently available gravitational-wave observatories, only the
low-frequency part (less than about 100 Hz) of gravitational waves
emitted in the early stage of binary inspirals of compact stellar
objects could be detected in the near future. However, as
suggested by Flanagan and Hinderer~\cite{Flanagan:08p021502}, 
such signals are still likely to be affected by the internal structure 
of the binary systems. For the same reason, it was proposed that the EOS of
nuclear matter could also be constrained from these gravitational-wave signals
\citep{Flanagan:08p021502,Hinderer:08p1216,Damour:09p084035,Lattimer_Love}.
In particular, the influence of stellar structure on the phase of
gravitational-wave signal in the  early stage of inspirals is
uniquely determined by  the quadrupole tidal  deformability
$\lambda_2$ defined by
\begin{equation}\label{lambda}
 Q_{ij} = -\lambda_{2}  {\cal E}_{ij}~,
\end{equation}
where $Q_{ij}$ is the traceless quadrupole moment tensor of the star 
and ${\cal E}_{ij}$ is the tidal field tensor inducing the deformation.
Besides, the term tidal Love number $k_2 \equiv 3 \lambda_2 /(2
R^5$) is also often referred to in the literature
\citep{Damour:09p084035,Yagi_14_PRD}. Numerous investigations have
been performed to study how the tidal deformability (or Love
number) depends on $R$, $M$, ${\cal C}$ and the EOS. In general,
obvious EOS dependence is observed if the tidal deformability (or
Love number) is considered as a function of  any one of $R$, $M$,
and ${\cal C}$
\citep{Hinderer:08p1216,Damour:09p084035,Lattimer_Love}.

Nonetheless, a set of almost universal relations, the $I$-Love-$Q$
relations, were discovered recently by Yagi and Yunes
\cite{Yagi:2013long,Yagi:2013}.  In such relations, three
dimensionless scaled physical quantities, namely, the scaled
moment of inertia ${\bar I} \equiv I/M^3$, the scaled quadrupole
tidal deformability ${\bar \lambda}_{2} \equiv \lambda_{2}/ M^5$,
and the scaled rotational quadrupole moment ${\bar Q} \equiv - M
Q/ J^2$, where $Q$ and $J$ are, respectively, the spin-induced
quadrupole moment and the angular moment of the star (see, e.g.,
\citep{Hartle:67p1005,Hartle:68:p907,Flanagan:08p021502,
Hinderer:08p1216,Damour:09p084035,Binnington:09p084018,Yagi:2013long,
Urbanec:13p1903} for methods to evaluate these quantities in
general relativity), are related through  three nearly
EOS-independent empirical formulas, which can be cast into the
following form \citep{Yagi:2013long,Yagi:2013}:
\begin{equation}
\ln y_i = a_i + b_i \ln x_i + c_i \left( \ln x_i \right)^2 + d_i
\left(\ln x_i \right)^3 + e_i \left( \ln x_i \right)^4 ,
\label{eq:ILoveQ_fit}
\end{equation}
where $(x_i,y_i)$ are any two of ${\bar I}$, ${\bar \lambda}_{2}$
and ${\bar Q}$,  $a_i$, $b_i$, $c_i$, $d_i$, and $e_i$ are some
fitting coefficients (see \citep{Yagi:2013long,Yagi:2013}). Thus,
given any one of these three scaled quantities, the remaining two
can be obtained. In addition, the $I$-Love-$Q$ relations also imply that the
degeneracy between the quadrupole moments and spins of NSs in
binary inspiral waveforms can be removed. Hence, with the advent
of second-generation gravitational-wave detectors, the averaged
(dimensionless) spin of binary systems could be measured with
accuracy up to $10\%$ \citep{Yagi:2013long,Yagi:2013}. Motivated
by the possible applications of the $I$-Love-$Q$ relations in
astrophysics, general relativity and fundamental physics, there
has recently been a surge of interest in this field
\citep{Yagi:2013long,Yagi:2013,Maselli:2013,Haskell_14_mnras,Doneva_14_apj,Pappas_14_prl,Chak_14_prl,Yagi_hair_GR,Stein_hair_apj,practical-no-hair}.

On the other hand, it is remarkable that ${\bar I}$ considered in
the $I$-Love-$Q$ relations is actually equal to the inverse square of
the effective compactness (i.e., $\eta^{-2}$) used in the $f$-$I$
relations (\ref{eq:fmode_real}) and (\ref{eq:fmode_img}). This
prompts us to study the possibility whether universal relations
directly linking ${\bar \lambda}_2$,  ${\bar Q}$ and
$\bar{\omega}_2$ exist. An additional question also arises
naturally. Could such relations exist in higher-order multipoles?
The focus of the present paper is to perform an in-depth
investigation on these issues. Specifically, we study the
so-called $f$-Love relation between  the $f$-mode frequency and the electric 
tidal deformability in the context of multipolar distortion in the following 
discussion.

\section{$f$-Love relations}
\label{f-Love}

\begin{figure*}[h!]
  \centering
    \includegraphics[scale=0.6]{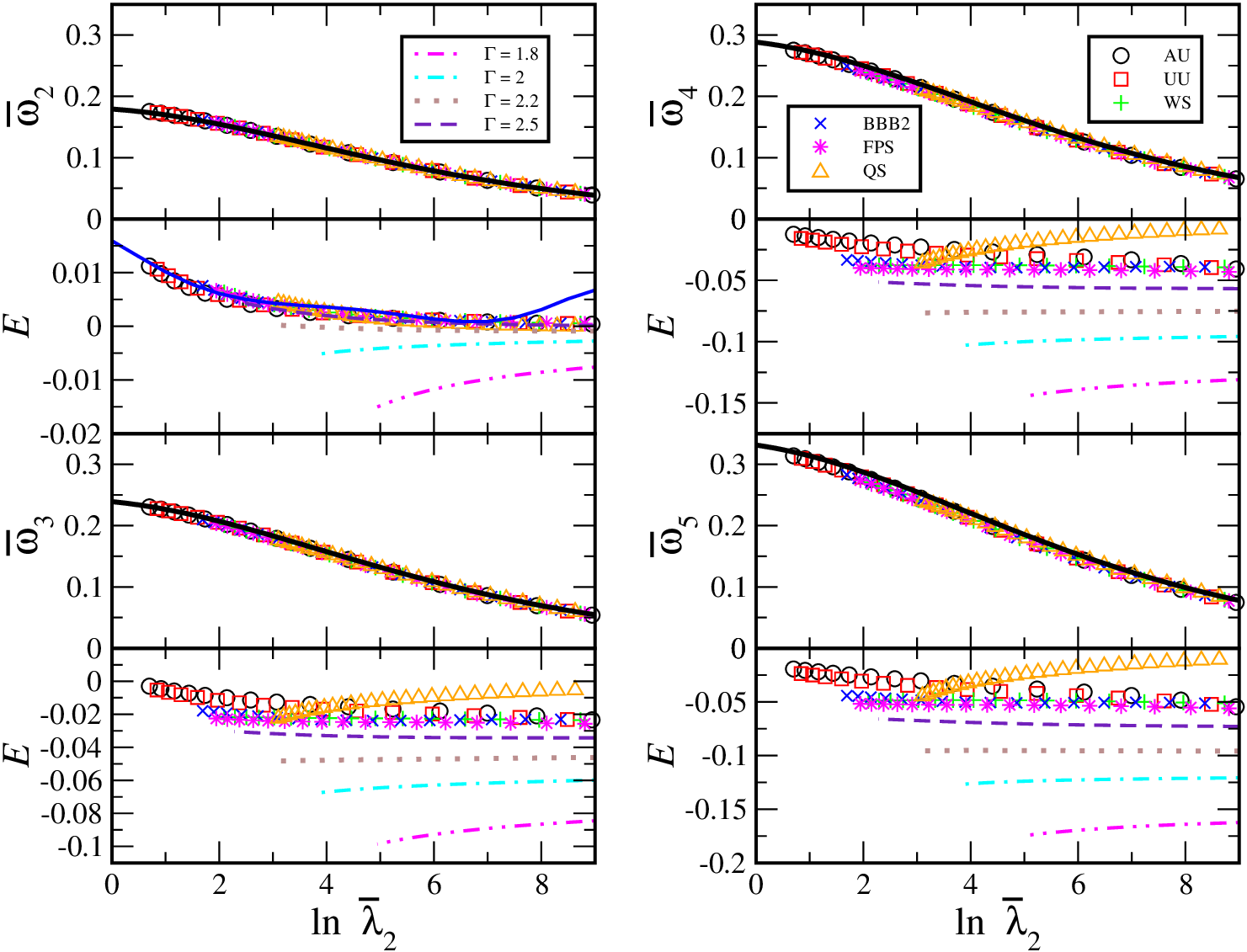}
  \caption{In the upper panel of each of the above four
  subfigures,  $ \bar{\omega}_l $ for  $l = 2$ (left top), 3 (left bottom), 4
  (right top), and $5$ (right bottom)
  is plotted against  $\ln \bar{\lambda}_{l'}$ for $l'=2$ case. Five kinds of realistic
NSs (including  AU, UU, WS, BBB2 and FPS EOSs), one QS, and the
incompressible star (indicated by the black solid line) are
considered (see the legend in the right-top subfigure). In the
  lower panel of each subfigure, the  fractional deviation $E$ from the incompressible limit
  (the  solid line in the upper panel)  is shown as a function of
$\ln \bar{\lambda}_{l'}$ for  $l'=2$ case.  In addition to the
data of the realistic stars mentioned above, we include in the
lower panel the data of polytropic stars with $\Gamma=
1.8,2,2.2,2.5$ (see the legend in the upper panel of the left-top
subfigure). For the diagonal case with $l=l'$, the (blue)
continuous line in the lower panel shows the fractional deviation
between the scaled frequencies given respectively by the best
fitting curve (\ref{best_eq}) and the incompressible limit. }
    \label{fig:k2_realistic}
\end{figure*}

\begin{figure*}[h!]
  \centering
    \includegraphics[scale=0.6]{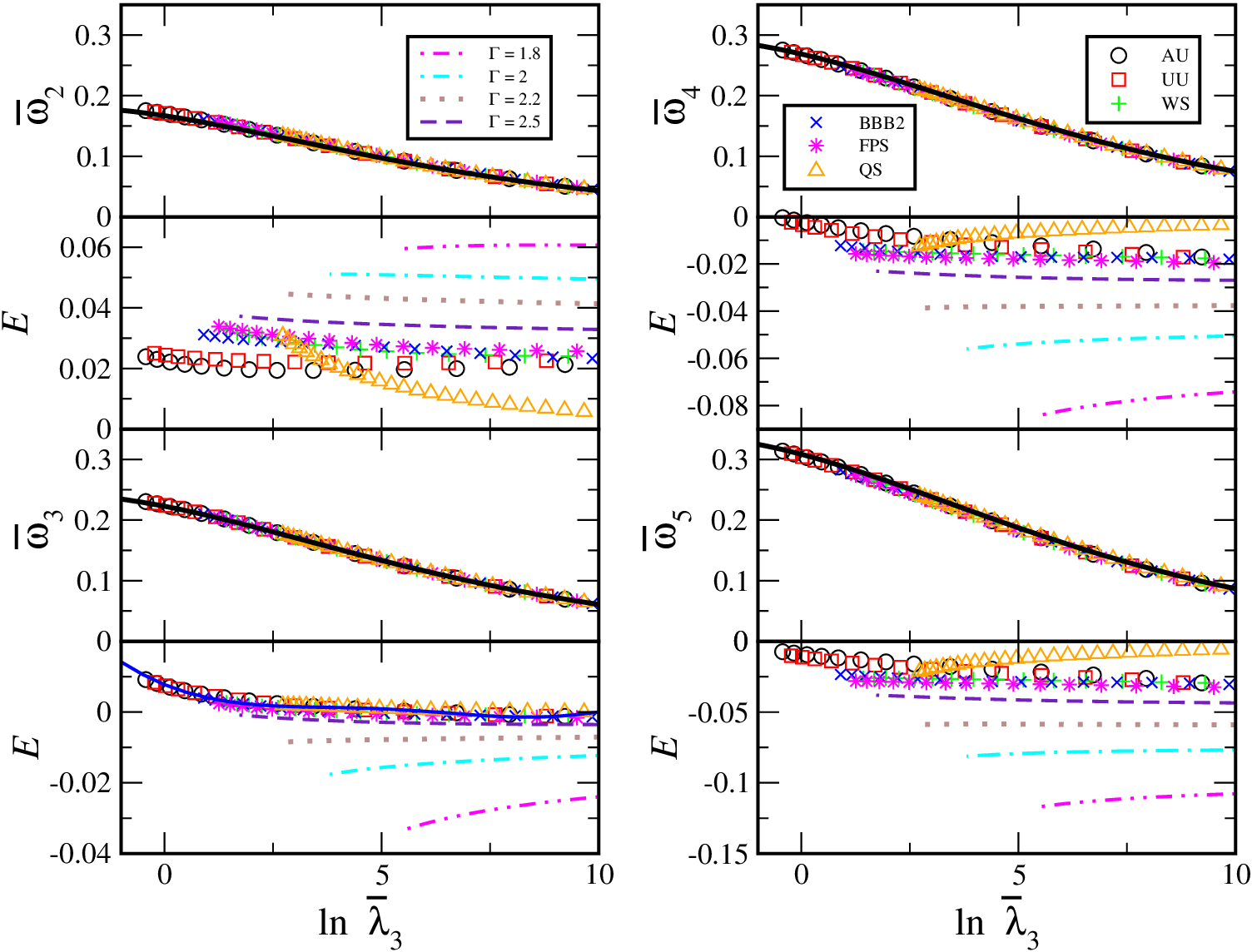}
  \caption{Similar to Fig.~\ref{fig:k2_realistic}, except that the case $l'=3$ is considered here. }
  \label{fig:k3_realistic}
\end{figure*}

\begin{figure*}[h!]
  \centering
    \includegraphics[scale=0.6]{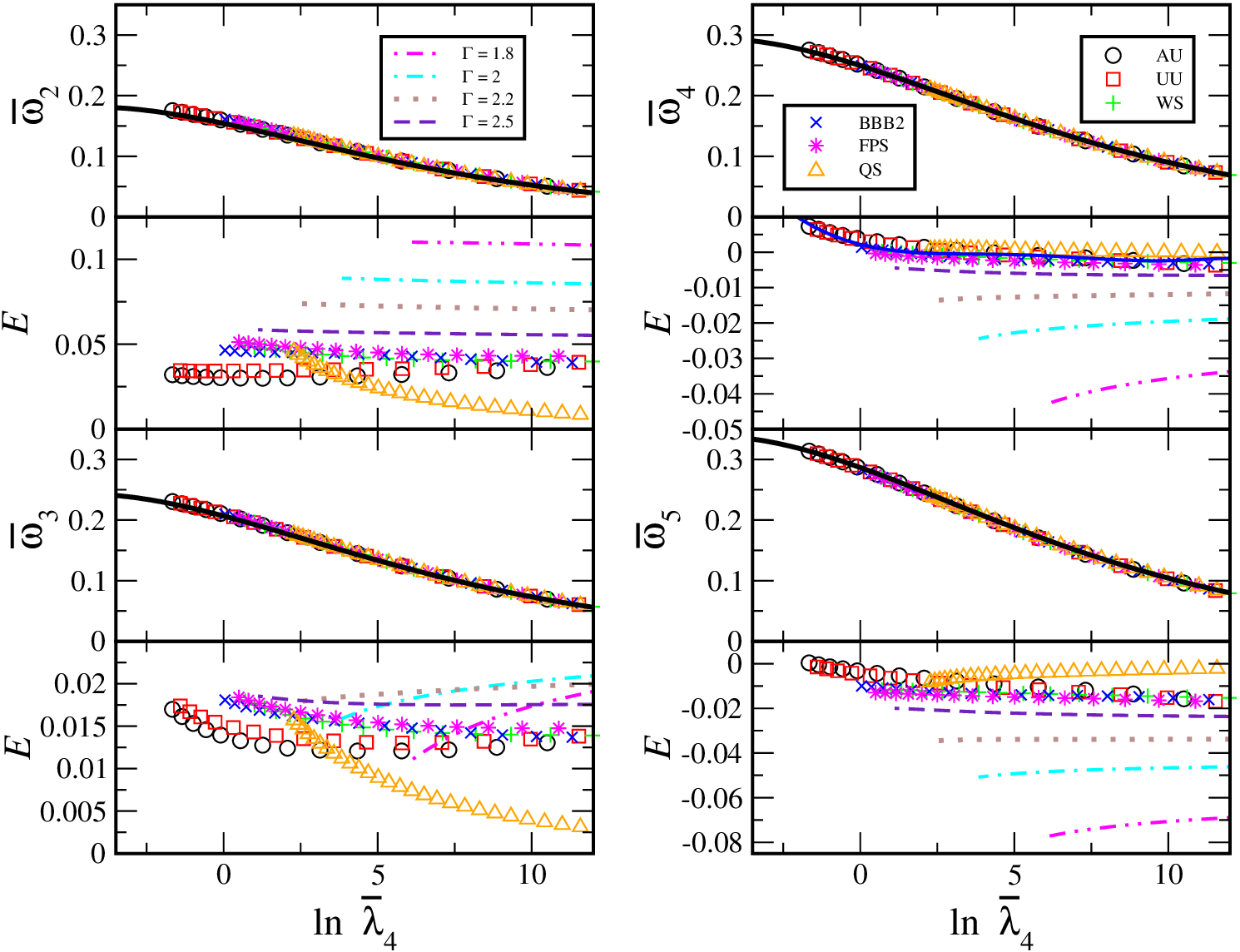}
\caption{Similar to Fig.~\ref{fig:k2_realistic}, except that the
case $l'=4$ is considered here.}
  \label{fig:k4_realistic}
\end{figure*}

\begin{figure*}[h!]
  \centering
    \includegraphics[scale=0.6]{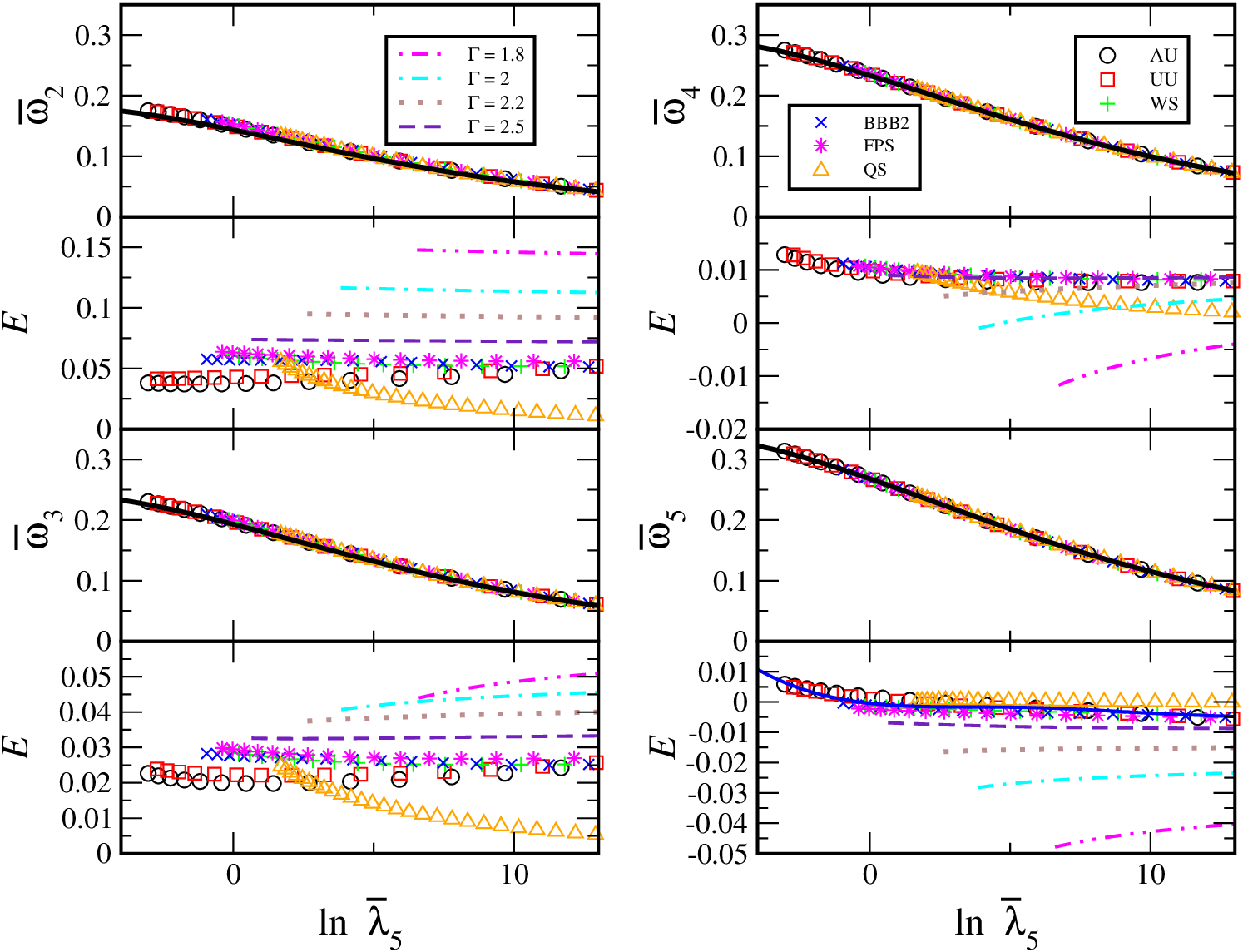}
  \caption{Similar to Fig.~\ref{fig:k2_realistic}, except that the case $l'=5$ is considered here. }
  \label{fig:k5_realistic}
\end{figure*}





\begin{figure*}[h!]
  \centering
    \includegraphics[scale=0.3]{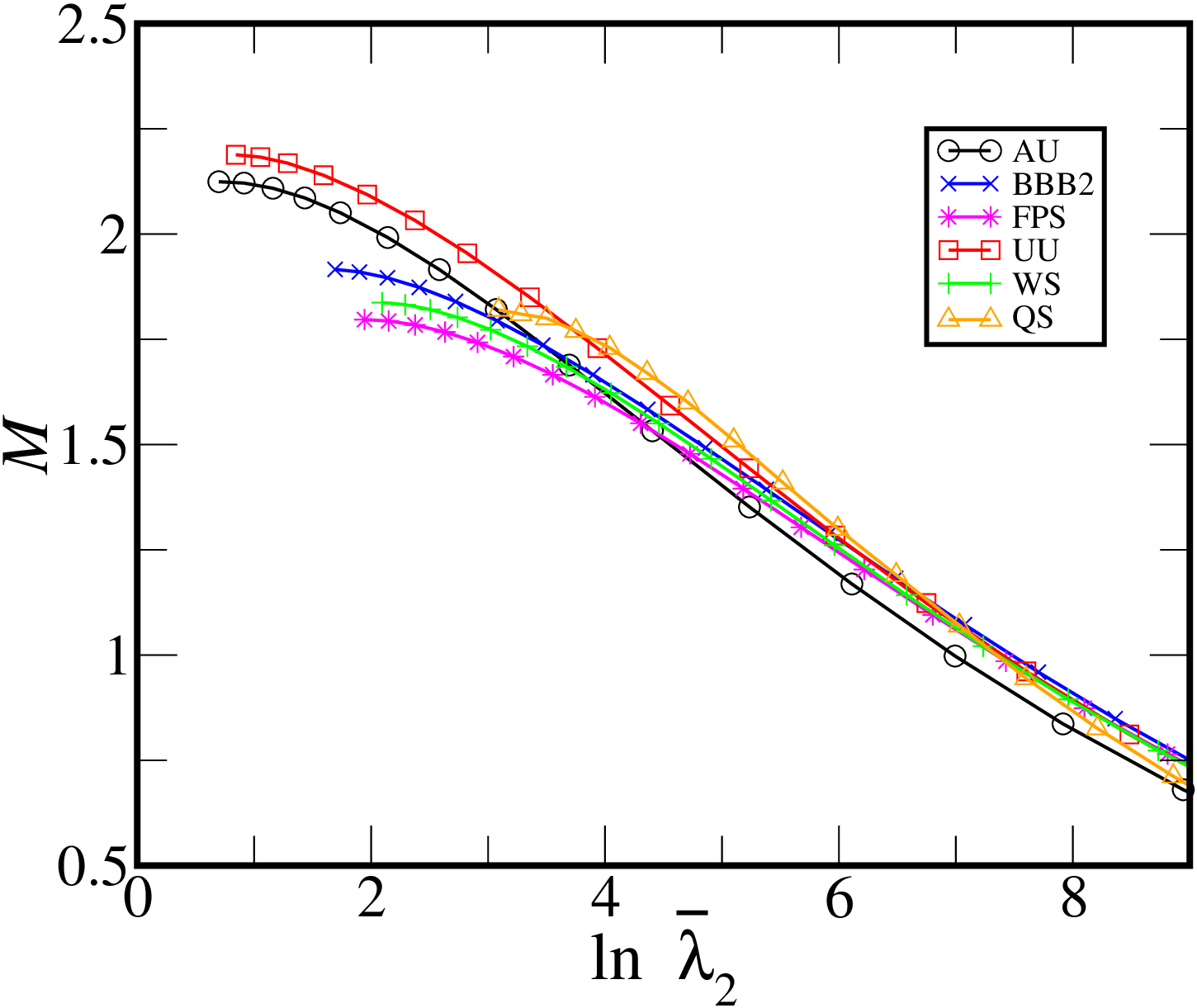}
  \caption{The masses (in $M_{\odot}$) of the realistic compact stars
considered in Figs.~\ref{fig:k2_realistic}-\ref{fig:k5_realistic}
are shown  against $\ln {\bar \lambda}_{2}$. }
  \label{m_k}
\end{figure*}

In the present paper,
we adopted the convention  introduced in
\citep{Damour:09p084035,Yagi_14_PRD} to discuss multipolar tidal
deformation. We consider the $l$th order electric tidal
deformability (see
\citep{PhysRevD.45.1017,Damour:09p084035,Yagi_14_PRD} for its
definition), $\lambda_l$, with $l$ being the angular momentum
index of the tidal field in consideration, as well as the
dimensionless $l$th order electric tidal deformability
\begin{equation}\label{deform}
    \bar{\lambda}_l \equiv \frac{\lambda_l}{M^{2l+1}},
\end{equation}
which is also related to $l$th order electric tidal Love number
$k_l$ through the relation \citep{Damour:09p084035}
\begin{equation}\label{Love}
 k_l    \equiv \frac{(2l-1)!!}{2}{\cal C}^{2l+1}\bar{\lambda}_l.
\end{equation}
As highlighted above, we investigate the relationship between the
$l$th multipole scaled $f$-mode frequency  $\bar{\omega}_l \equiv
M {\omega}_l$ and the $l'$th multipole dimensionless
deformability ${\bar \lambda}_{l'}$. In general, the values of $l$
and $l'$ can be distinct. In
Figs.~\ref{fig:k2_realistic}-\ref{fig:k5_realistic},
$\bar{\omega}_l$ is plotted against $\ln{\bar \lambda}_{l'}$ (the
upper panel), where $l=2,3,4,5$, and $l'=2$
(Fig.~\ref{fig:k2_realistic}), $l'=3$
(Fig.~\ref{fig:k3_realistic}), $l'=4$
(Fig.~\ref{fig:k4_realistic}) and $l'=5$
(Fig.~\ref{fig:k5_realistic}), for five kinds of realistic NSs
(respectively constructed with  AU \citep{WFF}, UU \citep{WFF}, WS
\citep{Lorenz:93p379,WFF}, BBB2 \citep{BBB2} and FPS
\cite{FPS,Lorenz:93p379} EOSs), one QS described by the MIT bag
model (see, e.g., \citep{Johnson,SS,Witten}), and incompressible
stars. For reference, the masses of the realistic compact stars
considered here are shown in Fig.~\ref{m_k} where $M$ (in
$M_{\odot}$) is plotted against $\ln {\bar \lambda}_{2}$.

At first sight, all such $\bar{\omega}_l$ versus $\ln{\bar
\lambda}_{l'}$ graphs demonstrate a certain degree of
EOS independence. In particular, all data points of realistic NSs
and QSs nearly coalesce onto the solid line representing the data
of incompressible stars. To further examine the degree of accuracy
of the universal relations manifested in these curves, we use the
case of incompressible stars as a reference and show the
fractional deviation $E \equiv (\bar{\omega}_l -
\bar{\omega}_l^{(0)})/\bar{\omega}_l^{(0)}$, where
$\bar{\omega}_l^{(0)}$ is the scaled frequency of incompressible
stars,  in the lower panel of each of these plots. It is clearly
shown that for a fixed $l'$, $|E|$ is the smallest for the
diagonal case where $l=l'$  and, in addition, also least sensitive
to EOS (including both NSs and QSs). In fact, $|E|$ is less than
0.01 in all diagonal cases considered here. Generally speaking,
for the off-diagonal cases where $l \ne l'$, the fractional
deviation $E$ of QSs deviates obviously from those of NSs.

To investigate how the fractional deviation $E$ in
$\bar{\omega}_l$ depends on the stiffness of a star, we also show
in Figs.~\ref{fig:k2_realistic}-\ref{fig:k5_realistic} $E$ versus
$\ln \bar{\lambda}_{l'} $ for polytropic stars with different
values of relativistic adiabatic index $\Gamma \equiv
[(\rho+p)/p](d p /d \rho)= 1+1/N$, where $\rho$, $p$, and $N$ are
energy density, pressure and the polytropic index, respectively.
Save for some ``atypical cases" which will be further discussed
later, the diagonal $f$-Love relations again yield the smallest
deviation from the incompressible limit and demonstrate the least
dependence on the adiabatic index $\Gamma$. We also note that
$|E|$ increases with the polytropic index $N$ and
 markedly grows  if $N>1$, especially for dense stars close to the
maximum mass limit (i.e., stars with small scaled tidal
deformability). As a rule of thumb, general relativistical effects
could apparently soften the stiffness of matter and lead to a
smaller effective value of the adiabatic index (see, e.g.,
\citep{Chand_1964ApJ_stab,Chand_1965ApJ_stab} and Chap. 6 of
\citep{BWN} for a heuristic explanation for the decrease in the
effective adiabatic index due to gravitational effect). For
example, NSs  with adiabatic index $\Gamma>$ 2 still become
unstable towards the high compactness end as a result of the
decrease in the effective value of $\Gamma$. Similarly,  the
reduction of the effective value of $\Gamma$ also explains the
increase in $E$ towards the maximum mass limit observed in
Figs.~\ref{fig:k2_realistic}-\ref{fig:k5_realistic}. The
above-mentioned observations clearly indicate that the stiffness
of nuclear EOS is a crucial factor affecting the $f$-Love relation
discovered here.

Moreover, some interesting behaviors of $E$ can be observed from
the graphs shown in
Figs.~\ref{fig:k2_realistic}-\ref{fig:k5_realistic}. For all
diagonal cases, $E$ is negative (except near the maximum mass
limit in Fig.~\ref{fig:k2_realistic}) and its magnitude decreases
with the stiffness (or $\Gamma$) of the EOS. For nondiagonal
cases showing $E$ in the plots of $\bar{\omega}_l$ versus
$\ln{\bar \lambda}_{l'}$, if $l>l'$, the above-mentioned behavior
of $E$ remains unchanged. In fact, $|E|$ also increases with $l$
for a fixed $l'$. However, the situation for cases with $l <l'$
becomes more complicated. For the diagrams of $E$ with $(l,l')=
(2,3), (2,4), (2,5), (3,5)$, $E$ becomes positive although its
magnitude still decreases with increasing stiffness (or increasing
$\Gamma$). On the other hand, for the ``atypical cases" with
$(l,l')= (3,4), (4,5)$ there are some crossings between lines
representing EOSs of different stiffness and the sign of $E$ is
not uniquely defined. Thus, the magnitude of $E$ is no longer a
good indicator for the stiffness of EOS.

Based on these observations, we figure out a phenomenological
method, whose analytical basis  will be provided in
Sec.~\ref{theory}, to explain the above-mentioned behavior of
$E$. Schematically we  expand $E$ in power series of $N$,
\begin{equation}\label{ES-1}
E(\ln{\bar \lambda}_{l'};l,l';N)=E_1(l,l')N+E_2(l,l')N^2,
\end{equation}
where the coefficients $E_1(l,l')$ and $E_2(l,l')$ in general
could also depend on ${\bar \lambda}_{l'}$, and only terms up to
$N^2$ are kept in the expansion. As shown in
Figs.~\ref{fig:k2_realistic}-\ref{fig:k5_realistic}, as long as
$N<1$, $E$ is insensitive to changes in $N$ for the diagonal case.
Hence, it is reasonable to assume that  if $l=l'$, $E_1 $ vanishes
and is
proportional to $l-l'$ (or any of its  positive powers). Without
loss of generality, we rewrite (\ref{ES-1}) as
\begin{equation}\label{ES-2}
E(\ln{\bar \lambda}_{l'};l,l';N)=-\alpha (l-l') N-\beta N^2,
\end{equation}
where $\alpha$ and $\beta$ (both are dependent on $l,l'$) are
assumed to be positive constants in order to explain the numerical
data shown in the figures. Therefore, for the diagonal cases with
$l=l'$, $E=-\beta N^2<0$ and $|E|$ increases with $N$. This
successfully captures the trend observed in
Figs.~\ref{fig:k2_realistic}-\ref{fig:k5_realistic}. In fact, we
have verified this point (within the accuracy of available
numerical data) in
Figs.~\ref{fig:k2_realistic}-\ref{fig:k5_realistic} in the
Newtonian limit (see also Table~\ref{tab:delta_square}).

On the other hand, for off-diagonal cases with $l>l'$, it follows
from (\ref{ES-2}) that $E$ is still negative and $|E|$ also
increases with $N$. Complication arises for off-diagonal cases
with $l<l'$. Several situations could be possible. Two of them
are: (i) If either $N$ is small or $|l-l'|$ is large, then the
first-order term dominates the second-order term. Consequently,
$E$ is positive and increases with $N$ [see the diagrams of $E$
with $(l,l')= (2,3), (2,4), (2,5), (3,5)$]. (ii) If the
first-order and the second-order terms are of similar magnitudes,
they could partly cancel each other out. Hence, $E$ is not of a
definite sign  and its magnitude may be rather small due to the
cancellation of these two terms. Such is the explanation for the
atypical cases mentioned above [cases with $(l,l')= (3,4),
(4,5)$]. Hence, in these atypical cases the magnitude of the
deviation could not used as a direct measure of the stiffness of
nuclear matter.

However, (\ref{ES-2}) has to be properly modified in order to
account for relativistic effects. As can be observed from the
diagonal cases in
Figs.~\ref{fig:k2_realistic}-\ref{fig:k5_realistic}, for a fixed
EOS $|E|$ usually increases towards the maximum mass limit, which
could be understood as the consequence of the general
relativistical softening effect mentioned above. Therefore, the
polytropic index $N$ used in (\ref{ES-2}) should be replaced by
its  modified value, which is expected to be larger than the
original value. Such replacement can qualitatively explain the
increase in $|E|$ observed for dense stars. Empirically, we find
that the modified polytropic index $N$ increases by an amount of
the order of the compactness of the star, which is in agreement
with the  post-Newtonian analysis carried out in
\citep{Chand_1964ApJ_stab,Chand_1965ApJ_stab}. On the other hand,
a subtlety can also be observed from these figures especially for
stiff stars in the small $\bar{\lambda}$ regime (i.e., dense
stars). It seems that an extra anomalous (positive) contribution
emerges in the right-hand side of (\ref{ES-1}). 
For example, in the diagonal case
of Fig.~\ref{fig:k2_realistic}, both the curves with
$\Gamma=2.5,2.2$ cross zero in the small $\bar{\lambda}$ regime.
We expect the magnitude of this extra term, which is more
important for stiff stars, grows with compactness and leads to
this anomalous behavior. However, the effect of this anomalous
term is quite small and we cannot extract its quantitative
dependence from our data.

To sum up the above observations, we find that both realistic NSs
and QSs obey the $f$-Love universal relation as shown in
Figs.~\ref{fig:k2_realistic}-\ref{fig:k5_realistic}. Such
universal behavior, which is shown to be insensitive to changes in
polytropic index $N$ as long as $N$ is less than unity (see
Figs.~\ref{fig:k2_realistic}-\ref{fig:k5_realistic}), is
attributable to the fact that realistic NSs far from the
theoretical minimum mass limit are made of stiff nuclear matter
with $N<1$. On the other hand, for QSs obeying the simple MIT bag
model (see, e.g., \citep{MITBM,ComStar,Weber-pulsar}), it is
straightforward to show that $N=\rho/(4B)-1$, where $B$ is the bag
constant. Unless for QSs with central density much higher than the
bag constant, $N$ is also less than unity. In fact, $N$ is almost
zero for low mass QSs. Hence, QSs also reveal similar $f$-Love
universal behavior.

Lastly, we note that the diagonal $f$-Love universal relation can be
represented by the following empirical formula
\begin{equation}\label{best_eq}
    \bar{\omega}_l = \bar{a}_0 + \bar{a}_1 x + \bar{a}_2 x^2 + \bar{a}_3 x^3 + \bar{a}_4 x^4~,
\end{equation}
where $x = \ln \bar{\lambda}_l$. The coefficients $\bar{a}_0$,
$\bar{a}_1$, $\bar{a}_2$, $\bar{a}_3$ and  $\bar{a}_4$ for
$l=2,3,4,5$ obtained from the best fit to the data of the
realistic stars considered in
Figs.~\ref{fig:k2_realistic}-\ref{fig:k5_realistic} are tabulated
in Table~\ref{best_table}. For reference, the fractional deviation
between the values of $\bar{\omega}_l$ given, respectively, by
(\ref{best_eq}) and the incompressible limit is also shown  in the
lower panel of each subfigure (the blue continuous line) of
Figs.~\ref{fig:k2_realistic}-\ref{fig:k5_realistic}. It can be
clearly seen that the fractional deviation is usually less than
0.01. This again supports our claim that the $f$-Love universal
relation of realistic stars follows that of incompressible stars.
\begin{widetext}
\begin{center}
\begin{table}[ht]
    \begin{tabular}{l|llll}
    \hline
    ~     & ~~~~~~$l = 2 $     & ~~~~~~$l = 3$ & ~~~~~~$l = 4$  & ~~~~~~$l = 5$    \\ \hline\hline
    $\bar{a}_0$ & ~~$1.820 \times 10^{-1}$  & $~~2.245 \times 10^{-1}$  & $~~2.501\times 10^{-1}$  & $~~2.681\times 10^{-1 }$\\
   $\bar{a}_1$  & $-6.836 \times 10^{-3}$ & $-1.500 \times 10^{-2}$ & $-1.646\times 10^{-2}$ &  $-1.638\times 10^{-2}$ \\
    $\bar{a}_2$ &$ -4.196 \times 10^{-3}$ & $-1.412\times 10^{-3}$  & $-5.897\times 10^{-4}$& $-2.497\times 10^{-4}$ \\
   $\bar{a}_3$  & $~~5.215 \times 10^{-4}$   & $~~1.832\times 10^{-4}$ & $~~8.695\times 10^{-5}$   & $~~4.712\times 10^{-5}$ \\
   $\bar{a}_4$  & $-1.857\times 10^{-5}$ & $-5.561\times 10^{-6}$ & $-2.368\times 10^{-6}$ & $-1.166\times 10^{-6}$ \\
\hline
    \end{tabular}
        \caption{The coefficients $\bar{a}_0$, $\bar{a}_1$, $\bar{a}_2$, $\bar{a}_3$ and  $\bar{a}_4$ in the empirical formula (\ref{best_eq})  are tabulated
    for
$l=2,3,4,5$.} \label{best_table}
\end{table}
\end{center}
\end{widetext}


\section{Newtonian analysis}
\label{theory}
\subsection{Generalized Tolman model}
In the previous section, we have shown that the diagonal $f$-Love
relation is particularly insensitive to variations in EOS and the
behavior of realistic stars and polytropic stars tends to that of
incompressible stars as long as these stars are sufficiently
stiff. On the other hand, the off-diagonal counterpart displays
more  sensitive dependence on stiffness. To provide these
observations a proper theoretical basis, at least in the Newtonian
limit, we consider a model compact star, referred to as the
generalized Tolman model (GTM) in the present paper, whose density
distribution $\rho(r)$ depends on the radial coordinate $r$ as
follows:
\begin{equation}\label{density}
\rho(r)=\rho_0 f(x,\delta),
\end{equation}
with $\rho_0$ being the central density, $x \equiv r/R$, $0 \le
\delta \le 1$ and
\begin{equation}\label{delta_model}
    f(x,\delta)\equiv 1- \delta x^2~.
\end{equation}
Here $\delta$ is a parameter
determining the stiffness of the EOS of the star, which can be
quantitatively measured by  a position-dependent effective
polytropic index $N_{\rm e}(\delta;x)$, where
\begin{eqnarray}\label{index}
 &&\left[N_{\rm e}(\delta;x)\right]^{-1} \equiv \frac{\rho}{p}\frac{dp}{d\rho}-1 \nonumber\\
 &=& \frac{(1 - \delta x^2)^2 (5 - 3\delta x^2)}{
 \delta [5 (1 - x^2) - 4 \delta (1 - x^4) +  \delta^2 (1 -
 x^6)]}
 -1.~~~~~
\end{eqnarray}

The GTM (\ref{delta_model}) reduces to nearly incompressible and
Tolman VII model stars in the limits $\delta \ll 1$ and $\delta
\approx 1$, respectively
\citep{Seidov_78,Seidov_04,Lattimer:2001,Lake_03_PRD,Lattimer_Love}.
For $\delta \ll 1$, it can be shown that $N_{\rm e} \approx
\delta$ for $x \leq \delta$. In fact,  near the center of the
star, (\ref{delta_model}) closely resembles the density
distribution of a polytropic star with polytropic index
$N=\delta$, whose density is,  to leading order in $N$, given by
\citep{Seidov_78,Seidov_04}
\begin{equation}\label{}
\rho(r)=\rho_0[ 1+N\ln(1-x^2)].
\end{equation}
Therefore, this model can nicely approximate nearly
incompressible stars in the small-$\delta$ limit. On the other
hand, if $\delta$ is equal to unity, the model obviously reduces
to the standard Tolman VII model, which has been shown to be a
good approximation of realistic NSs
\citep{Lattimer:2001,Lake_03_PRD,Lattimer_Love}. This is the
reason why the model (\ref{delta_model})  is coined here as the
generalized Tolman model. Besides, unless $\delta=1$, there is a
density discontinuity developed at the stellar surface, which
prevails in bare QSs (see, e.g., \citep{ComStar,Weber-pulsar} and
references therein). Hence, the GTM  is also capable of
reproducing such typical behavior of QSs.

Furthermore, we have also verified that the effective polytropic
index $N_{\rm e}$ in (\ref{index}) is always less than or equal to
unity for all physical choices of $\delta$ and $x$. For example,
for the case with $\delta=1$ (i.e., the standard Tolman VII
model), $N_{\rm e}=1$ at the stellar surface and decreases
monotonically to $2/3$ at the center $x=0$ \citep{HKLau}. For
$\delta<1$, the corresponding value of $N_{\rm e}$ is also less
than unity everywhere inside the star. Hence, we expect that GTM
is a valid model to emulate sufficiently stiff NSs and QSs whose
effective polytropic index is less than unity. In the following
discussion, we shall make use of the simplicity and manageability
of GTM to study the underlying physical mechanism of the universal
$f$-Love relation discovered here. First, two universal formulas
respectively connecting the $f$-mode frequency and the tidal
deformability to  mass moments of suitable orders are derived.
Each of these analytic formulas constitutes the generalization of
$f$-$I$ and $I$-Love relations to multipolar cases. The mass moment is
then eliminated from these two formulas and hence the $f$-Love
relation in the Newtonian limit is obtained.

\subsection{$f$-$I$ relation}
It can be shown from the variational method proposed by
Chandrasekhar~\cite{Chan1963} that the frequency of the $l$th multipole
$f$-mode oscillation,  $\omega_{l0} \equiv \omega_l   $ (the
subscript ``$0$" stands for $f$ mode and is suppressed unless
otherwise stated), of compact stars is approximately given by
\citep{tkchan}
\begin{equation}
\label{varfre}
\omega_l^2=\frac{2l(l-1)(2l-1)}{2l+1}\frac{\int_0^Rpr^{2l-2}{d}r}{\int_0^R\rho
r^{2l}{d}r}.
\end{equation}
Here $p$ is the pressure at radius $r$, which can be obtained from
the hydrodynamic equilibrium condition (see, e.g., \citep{cox}).
In (\ref{varfre}) the quasiincompressible fluid approximation
that the Lagrangian displacement $\boldsymbol{\xi}_{l0} \propto
{\boldsymbol \nabla}(r^l Y_{lm})$, where $Y_{lm}$ is the spherical
harmonic function, has been made. Such approximation stems from
the observation that in $f$-mode oscillations of stiff stars the
Lagrangian change in density is negligible except perhaps near the
stellar surface. As shown in Table~\ref{vartab}, the accuracy of
(\ref{varfre}) is excellent as long as the star concerned is
stiff. In particular, for incompressible stars (i.e.,  GTM stars
with $\delta=0$) the approximation turns out to be exact and for
QSs the percentage errors in $\omega_l^2$ ($l=2$ in the table) are
less than $4 \times
        10^{-3}\%$. On
the other hand, for polytropic stars with polytropic index $N$
less than unity and a realistic star (with FPS EOS) the percentage
error  is still less than $1.6\%$. However, the accuracy of the
approximation deteriorates with increasing value of the polytropic
index. As the polytropic index for typical NSs whose mass is
greater than $1 M_{\odot}$ is less than 1, the
quasiincompressible approximation is justified.

\begin{table}[ht]
        \centering
        \caption{$R^3\omega_2^2/(GM)$, where $\omega_2$ is the quadrupole $f$-mode oscillation frequency, is shown
         for various stars, including polytropic stars with
         different polytropic indices $N$, two GTM stars ($\delta=0,1$), a FPS NS (with central density
         $10^{15}\,\mathrm{g}\,\mathrm{cm}^{-3}$), and two QSs ($\cc=0.0811,0.1719$)
         with $B^{1/4}=154.5\,{\rm MeV}$.
        The four columns   indicate the EOS of the star, the exact values of $R^3\omega_2^2/M$,
         the corresponding approximate  values obtained
        from (\ref{varfre}) and the associated percentage errors, respectively.}
        \label{vartab}
        \begin{tabular}{l|lll}
        \hline
        \hline
         EOS&Exact& (\ref{varfre}) & Error\\  \hline
        Poly($N=3.0$)   &5.880& 10.60& 80.3\%\\
        Poly($ N=2.5)$&4.311&5.722& 32.7\%\\
        Poly($N=2.0$)  &3.026&3.431& 13.4\%\\
        Poly($N=1.7)$&2.441&2.635& 7.9\%\\
        Poly($N=1.5)$&2.119&2.221& 4.8\%\\
        Poly($N=1.0)$   &1.505&1.529& 1.6\%\\
        Poly($N=0.8)$&1.320&1.331& 0.8\%\\
        Poly($N=0.6)$&1.160&1.164& 0.3\%\\
        GTM($\delta=0$)      & 0.800 & 0.000& 0.0\%\\
        GTM($\delta=1$)& 1.324&1.333& 0.74\%\\
        FPS                   &1.397&1.402&0.4\%\\
         QS($\cc=0.0811)$        &0.8177  &0.8177& $4.4 \times
        10^{-4}\%$\\
        QS($\cc=0.1719)$        & 0.8403 & 0.8404& $3.9 \times
        10^{-3}\%$\\
        \hline \hline
        \end{tabular}
    \end{table}

For GTM, whose the density is given by (\ref{density}),
$\omega_l^2$ acquires the following explicit form:
\begin{eqnarray}\label{freq_sq}
\omega_l^2 = \frac{8 \pi \rho_0 G l(l-1)}{3} \frac{\int_0^1
x^{2l}f(x,\delta)f(x,3\delta/5){d}x}{\hat{f}\left(\frac{2l+1}{2l+3}
\delta \right)},
\end{eqnarray}
where $\hat{f}(y)=f(x=1,y)$. As expected, $\omega_l^2$ is
proportional to $\rho_0$. However, in general, the proportional
constant is dependent on the value of $\delta$, i.e., the EOS.

The GTM is characterized by two physical parameters, namely the
central density $\rho_0$  and the stellar radius $R$. Both the
$f$-mode oscillation frequency in (\ref{freq_sq}) and the tidal
deformability (to be discussed later) are dependent on them.
However, these two parameters are also expressible in terms of any
two of the mass moments:
\begin{eqnarray}\label{def_Moment}
  I_n \equiv \int_0^R \rho(r) r^{2+n} {d}r
  = \frac{\rho_0 R^{3+n}}{n+3}\hat{f}\left(\frac{n+3 }{n+5}\delta\right) ,
\end{eqnarray}
where $n=0,1,2,\cdots$, and, for simplicity, the solid angle
$4\pi$ is omitted in the definition of $I_n$, as
 follows:
 \begin{eqnarray}\label{moment}
\rho_0 &=& \frac{(n+3)^{3/n}
I_0^{1+3/n}\left[\hat{f}\left(\frac{n+3}{n+5}\delta\right)\right]^{3/n}}{3^{1+3/n}
I_n^{3/n}\left[\hat{f}\left(\frac{3\delta}{5}\right)\right]^{1+3/n}},
\\
R^n &=&\frac{(n+3)I_n \hat{f}\left(\frac{3\delta}{5}\right)}{3I_0
\hat{f} \left(\frac{n+3}{n+5}\delta\right)}.
\end{eqnarray}
Therefore, the $f$-mode oscillation frequency  and the tidal
deformability can be determined from $M=4 \pi I_0 $ and $I_n$
($n=1,2,3,\cdots$) and we have the freedom to choose the value of
$n$.

\begin{figure}[h!]
  \centering
    \includegraphics[scale=0.3]{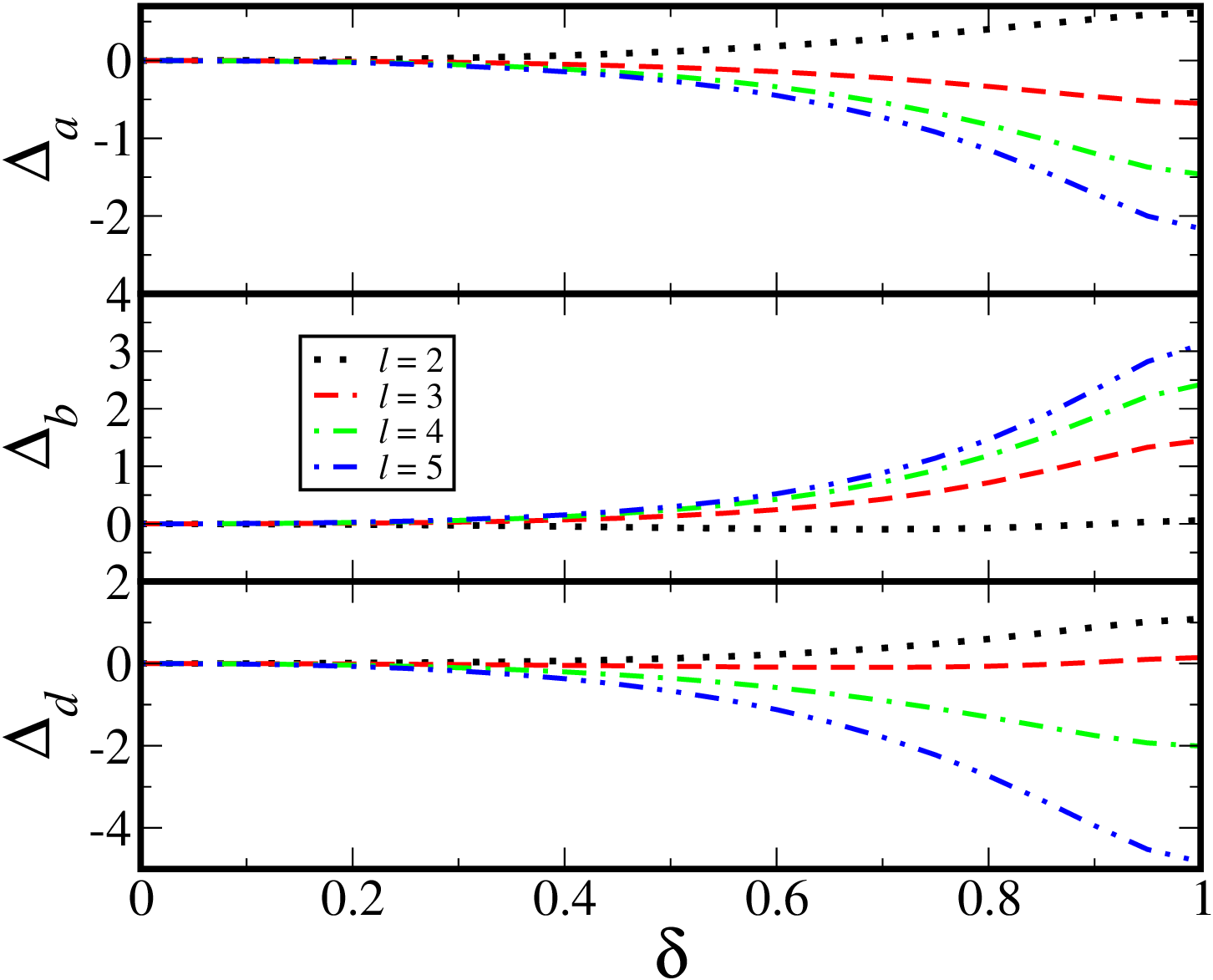}
   \caption{Plots of  $\Delta_a $ (top), $\Delta_b $ (middle) and $\Delta_d $
   (bottom) versus $\delta$  for GTM with $l  = 2, 3, 4$ and $5$. In all cases, $\Delta$'s are measured in percent.}
   \label{ABD}
\end{figure}

Substituting the above expression for $\rho_0$ into
(\ref{freq_sq}),
we can express $\omega^2_l$ in terms of simple algebraic functions
of $\delta$. In particular, apart from a trivial
$\delta$-independent proportional constant, we find that to
first order in $\delta$,
\begin{eqnarray}\label{freq_sq_expansion}
\omega_l^2 \propto \left(\frac{M^{n+3}}{I_n^{3}}\right)^{1/n}
\left\{1+\frac{6\delta(n+2-2l)}{5(n+5)(2l+3)} \right \},
\end{eqnarray}
and, most interestingly, the coefficient of $\delta$ in the above
expansion vanishes if $n=2 l-2$. Besides, it can also be shown
that $\delta=1$ is always a stationary point of $\omega_l^2$
irrespective of the values of $l$ and $n$. Therefore we expect
that the dependence of $\omega_l^2$ on $\delta$ is relatively weak
if $n=2 l-2$.  Hence, we arrive at the universal ``diagonal"
multipolar $f$-I relation
\begin{eqnarray}\label{freq_sq_expansion}
\bar{\omega}_l^2 = A_l(\delta)
\left(\frac{M^{2l-1}}{I_{2l-2}}\right)^{\frac{3}{2l-2}} ,
\end{eqnarray}
where the coefficient $A_l(\delta)$ is to first order independent
of $\delta$, and, in particular,
\begin{equation}
A_l(\delta=0)\equiv A_l^{(0)}=
\frac{2l(l-1)}{(2l+1)^{\frac{2l+1}{2l-2}}}\left(\frac{3}{4\pi}\right)^{\frac{3}{2l-2}}.
\end{equation}
It is interesting to note that to first order in $\delta$,
$(\bar{\omega}_2)^2 = A_2^{(0)}(M^{3}/I_{2})^{3/2}
\propto\eta^{3}$, which can be considered as the Newtonian
extension of the quadrupole $f$-$I$  relation (\ref{eq:fmode_real}).
In Fig.~\ref{ABD} (top panel), we show $\Delta_a \equiv
A_l(\delta)/A_l^{(0)}-1$ as a function of $\delta$. It is obvious
that $\Delta_a$ is limited to a few percent for all physical
values of $\delta$. Thus, the validity of the $f$-$I$ relation is
established.

\subsection{$I$-Love relation}
Similarly, we can evaluate the tidal deformability of GTM as
follows. In the Newtonian limit, the metric coefficient $H_l$
satisfies \citep{Damour:09p084035,Yagi:2013long,Yagi_14_PRD}
\begin{equation}\label{Poisson}
 \frac{d^2 H_l}{d r^2}   +\frac{2}{r}\frac{d  H_l}{d r}-\frac{l(l+1)}{r^2} H_l
 = -4\pi \rho\frac{d \rho}{d P} H_l ,
\end{equation}
which is essentially the Poisson equation in Newtonian gravity
with its right-hand side measuring the Eulerian change in density due to tidal
deformation, and the right-hand side of the above equation vanishes
identically outside the star. Once $ H_l$ is found,
$\bar{\lambda}_l$ is given by \citep{Damour:09p084035}
\begin{equation}\label{lambda}
\bar{\lambda}_l = \frac{1}{(2l-1)!!\,\cc^{2l+1}}
\frac{l-y_l}{l+1+y_l},
\end{equation}
where
\begin{eqnarray}
 y_l &\equiv& \left(\frac{r}{H_l}\frac{d H_l}{d r}\right)_{r=R^+}\nonumber \\
  &= &\left(\frac{r}{H_l}\frac{d H_l}{d r}-\frac{4 \pi R^3
\rho}{M}\right)_{r=R^-}.\label{bc}
\end{eqnarray}

For the GTM considered above, Eq.~(\ref{Poisson}) can be exactly
solved:
\begin{equation}\label{H}
H_l(r) = r^l \,_2F_1 \left(a,b;d;\zeta\right),
\end{equation}
where $_2F_1(a,b;d;\zeta)$ is the standard hypergeometric
function, with
\begin{eqnarray}
a&=&\frac{1+2l-\Xi}{4},\\
b&=&\frac{1+2l+\Xi}{4},\\
\Xi& = &\sqrt{4l^2+4l+41},\\
d&=&\frac{3}{2}+l,\\
\zeta&=&\frac{3 \delta r^2}{5R^2},
\end{eqnarray}
 and, as usual, regularity of $H_l$
at the origin is assumed. Hence, we can analytically find $y_l$:
\begin{eqnarray}\label{Y}
y_l = l-\frac{6\delta\, _2F_1 \left(a_1,b_1;d_1;\frac{3 \delta
}{5}\right)}{(2l+3)\, _2F_1 \left(a,b;d;\frac{3 \delta
}{5}\right)} -\frac{15(1-\delta)}{5-3\delta},
\end{eqnarray}
where $a_1=a+1$, $b_1=b+1$, and $d_1=d+1$. Expressing $R$ in terms
of $I_n$, we find that to first order in $\delta$,
\begin{eqnarray}\label{lambda_delta}
 &&(2l-1)!! M^{2l+1}\bar{\lambda}_l \left[\frac{(n+3)I_n
}{3I_0} \right]^{-\frac{2l+1}{n}}\nonumber \\
&=&
\left[\frac{3}{2l-2}+\frac{3(2l+1)(2l-2-n)\delta}{5(l-1)(2l+3)(n+5)}\right].
\end{eqnarray}
If $n=2l-2$, to first order in $\delta$, $\bar{\lambda}_l$ is
simply proportional to $(I_n /M)^{(2l+1)/n}$ with a
$\delta$-independent proportional constant. Such a case also holds
if $1-\delta \ll 1$ for all values of $n$. The situation is
similar to the analysis developed previously for the $f$-$I$ relation.
As a consequence, the universal ``diagonal" multipolar $I$-Love
formula can be expressed as
\begin{equation}\label{I-Love}
\bar{\lambda}_l = B_l \left(\frac{I_{2l-2} }{M^{2l-1}}
\right)^{\frac{2l+1}{2l-2}},
\end{equation}
where $B_l(\delta)$ depends only weakly on $\delta$, with
$(dB_l/d\delta)_{\delta=0}=0$ and
\begin{eqnarray}\label{B}
&&B_l (\delta=0)\equiv B_l^{(0)}\nonumber
\\&=&\frac{3}{2(l-1)[(2l-1)!!]}\left[\frac{4 \pi(2l+1) }{3}
\right]^{\frac{2l+1}{2l-2}}.
\end{eqnarray}
For $l=2$, Eq.~(\ref{I-Love}) leads to $\bar{\lambda}_2 =
B_2^{(0)} (I_{2} /M^{3})^{5/2}$ to first order in $\delta$, which
is in agreement with the result obtained in \citep{Yagi:2013long}.
Figure~\ref{ABD} (middle panel) plots $\Delta_b \equiv
B_l(\delta)/B_l^{(0)}-1$ against $\delta$ for $l=2,3,4,5$. Again
$\Delta_b$ is at most a few percent for $0 \le \delta \le 1$,
which guarantees the accuracy of the multipolar $I$-Love relation.
However, it is obvious that EOS dependence of such diagonal
relations grows gradually with increasing value of $l$.

\subsection{ $f$-Love relation}
\begin{figure}[h!]
  \centering
    \includegraphics[scale=0.3]{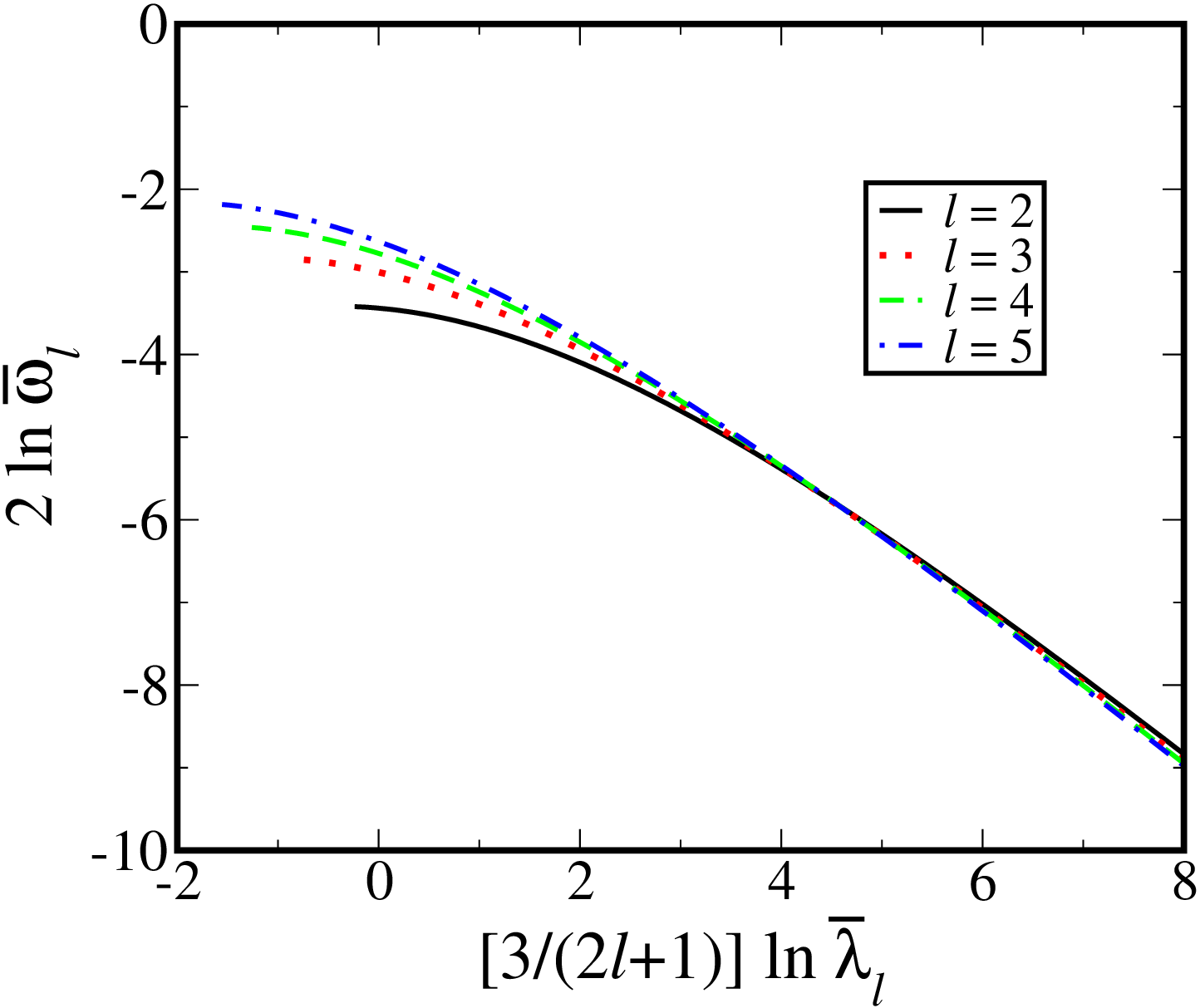}
   \caption{Plot of   $2 \ln \bar{\omega}_{l}$ against
    $[3/(2l+1)]\ln \bar{\lambda}_l$ for incompressible stars with $l  = 2, 3, 4$ and $5$.}
   \label{fig:scaled}
\end{figure}

Eliminating the mass moment $I_{2l-2}$ from
(\ref{freq_sq_expansion}) and (\ref{I-Love}), we  arrive at the
Newtonian form of the diagonal multipolar $f$-Love universal
relation:
\begin{equation}\label{f-Love-1}
    \bar{\omega}_l^2 \bar{\lambda}_l^{\frac{3}{2l+1}} = D_l(\delta),
\end{equation}
where $D_l(\delta)$ is an EOS-insensitive function with
$(dD_l/d\delta)_{\delta=0}=0$ and for $\delta=0$:
\begin{eqnarray}\label{D}
&&D_{l}  (\delta=0)\equiv D_l^{(0)}\nonumber \\&=&
\frac{2l(l-1)}{2l+1}\left[\frac{3}{2(l-1)(2l-1)!!}\right]^{\frac{3}{2l+1}}.
\end{eqnarray}
The plot of $\Delta_d \equiv D_l(\delta)/D_l^{(0)}-1$ versus
$\delta$ in Fig.~\ref{ABD} (bottom panel) firmly corroborates the
EOS-insensitive multipolar $f$-Love relation derived above. Besides,
the universal Newtonian  relation in (\ref{f-Love-1}) is verified
in Fig.~\ref{fig:scaled} where $2 \ln \bar{\omega}_{l}$ is plotted
against   $[3/(2l+1)]\ln \bar{\lambda}_l$ with $l=2,3,4,5$ for
incompressible stars (i.e., $\delta =0$). In the Newtonian limit,
all curves tend to straight lines with the same slope but slightly
different intercepts, which agree with the value given by
(\ref{D}).

Similarly, the off-diagonal $f$-Love relation can also be obtained
and is shown below for purpose of comparison:
\begin{eqnarray}\label{off-1}
\bar{\omega_l}^2\bar{\lambda}_{l'}^{\frac{3}{2l'+1}} =
D_{ll'}^{(0)}\left[1-\frac{12(l-l')\delta}{5(2l'+3)(2l+3)}+...\right],
\end{eqnarray}
where
\begin{equation}\label{off-2}
D_{ll'}^{(0)}=\frac{2l(l-1)}{2l+1}
\left[\frac{3}{2(l'-1)(2l'-1)!!}\right]^{\frac{3}{2l'+1}}.
\end{equation}
In general, for $l \ne l'$, the $\delta$-term in the right-hand side of
(\ref{off-1}) does not vanish, implying that the off-diagonal
universal relation would demonstrate more obvious EOS dependence.
This clearly  explains  our findings in the previous section and
further supports the phenomenological model proposed in
(\ref{ES-1}) and (\ref{ES-2}).

Moreover, we have evaluated $\bar{\omega}_l$, $\bar{\lambda}_l$
and $I_{2l-2}$ ($l=2,3,4,5$) for polytropic stars with different
polytropic indices and evaluated $A_l$, $B_l$ and  $D_l$ from
(\ref{freq_sq_expansion}), (\ref{I-Love}) and (\ref{f-Love-1}),
respectively. In Table~\ref{tab:delta_square} we compare these
values with their incompressible counterparts to find $\Delta_a$,
$\Delta_b$ and $\Delta_c$. Again we find that the deviations are
small as long as $N \le 1$ and indeed proportional to $N^2$ (or
higher) within the accuracy our numerical data. In fact, for the
$l=2$ case all $\Delta$'s approximately  behave as $N^4$, while in
other cases they  grow gradually with $l$.

\begin{widetext}
\begin{center}
\begin{table}
    \begin{tabular}{l|llll}
    \hline
     $N$     & $l=2$                        & $l=3$                 & $l=4$                 & $l=5$                 \\ \hline\hline
    $5/3$  & $(-4.6,3.3,-2.7)$        & $(-11,9.9,-7.6)$  & $(-15,14,-11)$    & $(-16,16,-13)$    \\
    $1.25$ & $(-1.5,1.1,-0.93)$       & $(-5.4,4.7,-3.5)$ & $(-7.6,6.9,-5.5)$ & $(-8.8,8.4,-6.8)$ \\
    $1$    & $(-0.61,0.34,-0.4)$      & $(-3.1,2.7,-2.0)$  & $(-4.6,4.1,-3.3)$ & $(-5.5,5.1,-4.2)$ \\
    $5/6$  & $(-0.24,0.11,-0.17)$     & $(-2.0,1.7,-1.2)$  & $(-3.0,2.8,-2.1)$  & $(-3.7,3.5,-2.8)$ \\
    $2/3$  & $(-0.011,-0.022,-0.024)$ & $(-1.1,1.0,-0.68)$ & $(-1.8,1.8,-1.3)$ & $(-2.3,2.2,-1.7)$ \\ \hline
    \end{tabular}
    \caption{In each entry the triplet ($\Delta_a\times 100\%$, $\Delta_b\times 100\%$, $\Delta_c\times 100\%$) is shown
    for various values of the polytropic index $N$ and $l=2,3,4,5$. }
    \label{tab:delta_square}
\end{table}
\end{center}
\end{widetext}

\section{$f$-$I$-Love relation}
\label{f-I-Love} Instead of eliminating the mass moment
$I_{2l-2}$, we can remove  mass $M$ from (\ref{freq_sq_expansion})
and (\ref{I-Love}) to obtain an equation linking ${\lambda_l}$,
${\omega}_l^2$ and ${I_{2l-2}}$ together:
\begin{equation}\label{f-Love-moment-dim}
 {\lambda_l} {\omega}_l^2 =\frac{4\pi l }{(2l-1)!!}
{I_{2l-2}}.
\end{equation}
This equation, coined as the $f$-$I$-Love relation in the present
paper, indeed connects three physical quantities, namely the
$f$-mode frequency $\omega_l$, tidal deformation $\lambda_l$ and
the mass moment $I_{2l-2}$ together. It is noteworthy that each of
them carries proper dimensions and the mass $M$ completely
disappears in (\ref{f-Love-moment-dim}). Comparing the $f$-$I$-Love
relation (\ref{f-Love-moment-dim}) with the $f$-Love relation
(\ref{f-Love-1}), the mass moment $I_{2l-2}$ in the former
actually plays the role of $M$ in the latter.

We have derived the $f$-$I$-Love relation (\ref{f-Love-moment-dim})
from the GTM assumed above. However, to show the robustness of
(\ref{f-Love-moment-dim}), in the following we apply the linear
response theory (Green's function method) to quasiincompressible
stars to develop an independent proof for it.

In Newtonian theory, the steady state response of a star to an
external time-dependent potential $U({\bf r})\exp(-i \omega t) $
of frequency $\omega$ can be obtained from the Green's function
method to be detailed as follows \citep{Chan1963,Press_77_ApJ}.
First of all, the normal modes of a star are defined by the
solutions to the eigenvalue equation:
\begin{equation}\label{eigen}
({\cal L} - \rho_{\rm e} \omega^2_{ln}) \boldsymbol{\xi}_{ln}= 0,
\end{equation}
where $\rho_{\rm e}(r)$ is the density distribution of the
equilibrium state and $\omega_{ln}$ and $\boldsymbol{\xi}_{ln}$ are
the eigenfrequency and the Lagrangian displacement of the $n$th
($n=0,1,2,\ldots$) oscillation mode carrying angular momentum
$l=0,1,2,3,\ldots$. The $l=0$ case corresponds to radial
oscillation, $l=1$ case usually represents translational motion,
and for $l \geq 2$ cases the star undergoes nonradial
oscillations. Unless otherwise stated explicitly, the
$z$ component of angular momentum, $m$, is suppressed in our
equations. We adopt the convention where $\omega_{l0} <\omega_{l1}
< \omega_{l2} \cdots$. For barotropic stars, which is always
assumed in our analysis, the 0th mode is the $f$ mode. Besides,
$-{\cal L} \boldsymbol{\xi}$ in general represents the net
internal restoring force (including pressure force and the
gravitational force due to the star itself) acting on a fluid
element, which can be obtained from the Lagrangian displacement
$\boldsymbol{\xi}$. The explicit form of ${\cal L}$ can be found
in \citep{Chan1963}. Most importantly, ${\cal L}$ is a Hermitian
operator guaranteeing the orthogonality relation of
$\boldsymbol{\xi}_{ln}$ \citep{Chan1963}:
\begin{equation}\label{ortho}
\langle \boldsymbol{\xi}_{lj}^* | \boldsymbol{\xi}_{l'n} \rangle
\equiv \int_V \rho_{\rm e} \boldsymbol{\xi}_{lj}^* \cdot
\boldsymbol{\xi}_{l'n} d^3 r = \delta_{ll'}\delta_{jn},
\end{equation}
where $V$ here denotes the volume of the star and the displacement
fields are properly normalized to accord with (\ref{ortho}).

In the presence of an external time-dependent gravitational
potential $U({\bf r})\exp(-i \omega t) $, the steady state linear
response of the Lagrangian displacement $\boldsymbol{\xi}$ is
given by solution of the inhomogeneous equation
\begin{equation}\label{steady}
({\cal L} - \rho \omega^2) \boldsymbol{\xi} = -  \rho_{\rm e}
{\boldsymbol \nabla} U.
\end{equation}
The right-hand side of the above equation is actually the external
gravitational force exerted on the star. Expanding
$\boldsymbol{\xi}$ in terms of the complete set of
$\boldsymbol{\xi}_{ln}$ and using (\ref{eigen})-(\ref{steady}), we find that
\begin{equation}\label{response}
 \boldsymbol{\xi}   = \sum_{l,n} \frac{\langle \boldsymbol{\xi}_{ln} |  {\boldsymbol \nabla} U\rangle
 \boldsymbol{\xi}_{ln}
 }{\omega^2-\omega_{ln}^2}= \sum_{l,n} B_{ln} \boldsymbol{\xi}_{ln}.
\end{equation}

In particular, if the external potential is a time-independent
tidal potential in the $l$th ($l \ge 2$) multipolar sector,
namely $U_{\rm e} = r^l Y_{lm} (\theta,\phi)$, where $Y_{lm}
(\theta,\phi)$ is the standard spherical harmonics of angles
$\theta$ and $\phi$, a corresponding multipole moment defined by
\begin{equation}\label{moment}
Q_l =  \frac{4 \pi}{2l+1}\int_V r^l \delta\rho Y_{lm}^*
(\theta,\phi) d^3 r
\end{equation}
with $\delta\rho(\boldsymbol{r})$ being the Eulerian change in
density, is induced. The induced multipole moment in turn sets an
additional potential $U_{\rm p} =Q_l  Y_{lm} (\theta,\phi)
/r^{l+1}$ outside the star.

In the linear response regime $\delta\rho(\boldsymbol{r})=
-{\boldsymbol \nabla} \cdot (\rho_{\rm e} \boldsymbol{\xi})$. The
expansion in (\ref{response}) then readily leads to
\begin{eqnarray}\label{moment2}
Q_l& =& \frac{4 \pi}{2l+1} \sum_{l',n}  B_{l'n} \langle
{\boldsymbol \nabla}(r^l
Y_{lm})| \boldsymbol{\xi}_{l'n} \rangle \nonumber  \\
 &=& \frac{- 4 \pi}{2l+1} \sum_{l',n} \frac{
  |\langle  {\boldsymbol \nabla}(r^l
Y_{lm})| \boldsymbol{\xi}_{l'n} \rangle |^2
 }{\omega_{l'n}^2}
\end{eqnarray}

So far the calculation has been exact up to the leading order of
the perturbing potential. For $f$-mode oscillations of
quasiincompressible stars, the Lagrangian change in the density
vanishes and hence ${\boldsymbol \nabla} \cdot
\boldsymbol{\xi}_{l0}=0$. Taking into consideration the fact that
$\boldsymbol{\xi}$ is derivable from a scalar potential (see,
e.g., \citep{cox}), we arrive at the approximate formula
$\boldsymbol{\xi}_{l0}= {\boldsymbol \nabla}(r^l Y_{lm})/\langle
{\boldsymbol \nabla}(r^l Y_{lm})|{\boldsymbol \nabla}(r^l Y_{lm})
\rangle^{1/2}$. We note that such approximation has been put
forward by Chandrasekhar~\cite{Chan1963} as the trial input of a variational
principle, which is used in Sec.~\ref{f-Love} to evaluate
$f$-mode oscillation frequencies. By orthogonality of the normal
modes, only the $f$ mode with the same angular momentum index $l$
has to be included in the sum in (\ref{moment2}), and after
integrating by parts, we show that
\begin{equation}\label{Q-f-I}
Q_l \omega_{l0}^2 = -4 \pi l I_{2l-2} .
\end{equation}
By virtue of (\ref{Love}), (\ref{Q-f-I}) and the definition of
tidal Love number \citep{Damour:09p084035,Yagi_14_PRD},
\begin{equation}\label{Love-moment}
k_l = - \frac{Q_l}{2R^{2l+1}}=-\left(\frac{U_{\rm p}}{2U_{\rm
e}}\right)_{r=R},
\end{equation}
the general $f$-$I$-Love relation involving three dimensionless
quantities $ M \omega_{l0} \equiv \bar{\omega}_l$,
$I_{2l-2}/M^{2l-1}$ and $\bar{\lambda_l}$
\begin{equation}\label{f-Love-moment}
 \bar{\lambda_l} \bar{\omega}_l^2 = \frac{4\pi l }{(2l-1)!!}\left(\frac{I_{2l-2}}{M^{2l-1}}\right)
\end{equation}
is established, which holds for sufficiently stiff stars and
$l=2,3,4,\ldots$.

It is remarkable that (\ref{f-Love-moment-dim}) and
(\ref{f-Love-moment}) are in fact equivalent to each other. While
the former is a consequence of $f$-$I$ and $I$-Love relations, the
latter is derived directly from the Green's function method. Given
the $f$-$I$-Love relation (\ref{f-Love-moment}) obtained from the
linear response theory, the two seemingly
independent $f$-$I$ and $I$-Love relations are in fact the consequence
of each other for quasiincompressible stars characterized by
stiff EOSs. Hence, the interrelationship between the $f$-$I$ and
$I$-Love relations is clearly shown from the $f$-$I$-Love relation.

\section{Conclusion and discussion}
\label{sec:conclude} Although the $f$-$I$ relation and the $I$-Love-$Q$
relations have recently been discovered in the quadrupolar sector
and various potential applications of them have been proposed
\citep{Lau:2010p1234,Yagi:2013long,Yagi:2013}, the reasons for the
validity of these two relations and their interrelationship are
not yet fully understood. Yagi and Yunes~\cite{Yagi:2013long,Yagi:2013}
suggested two possible reasons for the $I$-Love-$Q$ relations: (i) the
relations are most sensitive to the stellar matter in an outer
layer between $70 \%$ and $90 \%$  of  the radius of a NS and the
EOS there is quite unified; and (ii) the internal stellar
structure of NSs is gradually  effaced as the black-hole limit is
approached and hence NSs reveal similar behavior. On the other
hand, more recently Yagi {\it et al.}~\cite{whyI} 
found that the $I$-Love-$Q$ relations
are actually dominated by the outer core lying in a region bounded
between $50 \%$ and $90 \%$ of the stellar radius. However, given
this finding, the $I$-Love-$Q$ relations can no longer be attributed
to the similarity of EOSs in the low-density regime. In addition,
both $f$-$I$ relation and the $I$-Love-$Q$ relations are valid for QSs
\citep{Lau:2010p1234,Yagi_14_PRD}, whose EOS and stellar structure
are completely different from those of NSs especially in the outer
layer. Formally speaking, the outer layer of bare QSs can be
considered as incompressible, while that of NSs is rather soft
with an adiabatic index of about 1.4.

In the present paper we perform an in-depth examination on the
relationship between the $f$-$I$ relation and the $I$-Love-$Q$ relations,
in turn propose a robust multipolar $f$-Love relation, and study the
physical origin of these universal relations for compact stars in multipolar 
distortions. The multipolar $f$-Love relation discovered here is generally 
valid in any angular momentum sector with $l \ge 2$ (albeit EOS dependence 
increases with $l$) and for realistic compact stars (including both NSs and 
QSs) constructed with
different prevailing nuclear EOSs. We pinpoint that such universal
behavior of realistic stars indeed follows closely that of
incompressible stars, which are chosen as the standard to
benchmark other stars. As shown in
Figs.~\ref{fig:k2_realistic}-\ref{fig:k5_realistic}, as long as
the polytropic index $N$ of a star is not greater than 1, the
fractional deviation in $f$-mode frequency, as compared with the
incompressible counterpart, is less than $2\%$ and is inert to
changes in $N$. Therefore, we claim here that the stiffness of
nuclear matter at large densities is the crux of these multipolar
universal relations.

Through the GTM, which is able to mimic realistic stars with $N
\le 1$, we carry out detailed Newtonian analysis to show that both
$\bar{\omega}_l$ and $\bar{\lambda}_l$ are related to $I_{2l-2}$,
the $(2l-2)$th mass moment, in a way insensitive to changes in
the parameter $\delta$. Hence, after eliminating $I_{2l-2}$ from
the $f$-$I$ and $I$-Love relations, the Newtonian form of the diagonal
multipolar $f$-Love relation (\ref{f-Love-1}) is readily
established, which provides a strong support to the relativistic
$f$-Love relation discovered here. On the other hand, we also use
the linear response theory to derive a $f$-$I$-Love relation for
Newtonian stars with high stiffness (say, $N \le 1$). Any two of
the $f$-$I$, $I$-Love and $f$-$I$-Love relations will imply the validity of
the other one. More interestingly, in the $f$-Love relation
(\ref{f-Love-1}), the $f$-mode frequency and the tidal
deformability are related with the mass as a parameter in the
formula. However, in the $f$-$I$-Love relation
(\ref{f-Love-moment-dim}) the $f$-mode frequency, the tidal
deformability and the $(2l-2)$th mass moment are directly linked
together in the absence of the knowledge of the mass.

In general, for NSs far from their upper and lower mass limits,
they are stiff enough to guarantee the universal formulas studied
here. However, deviations from these universal formulas for stars
with masses close to either of these two mass limits could arise
for the following reasons. Near the maximum mass limit, the strong
gravitational attraction effectively softens the nuclear matter
and hence the star can no longer be approximated by an
quasiincompressible star. From
Figs.~\ref{fig:k2_realistic}-\ref{fig:k5_realistic}, we can see
that the magnitude of $E$ gradually grows larger and displays
stronger dependence on the polytropic index as the star concerned
approaches the maximum mass limit. In fact, such a softening
mechanism  seems to reduce the adiabatic index by an amount of the
order of the compactness of the star
\citep{Chand_1964ApJ_stab,Chand_1965ApJ_stab} and hence, following
directly from (\ref{ES-2}), leads to larger deviation from the
incompressible star. Notwithstanding this, the universal formulas
still hold around the maximum mass limit because the softening
effect will at the same time destabilize the
star \citep{Chand_1964ApJ_stab,Chand_1965ApJ_stab}. As a result,
the star becomes unstable before it could further deviate from the
universal formulas. On the other hand, near the low mass limit,
the star is primarily made of soft nuclear matter with adiabatic
index $\Gamma \approx 1.4 $ and therefore deviations from the
behavior of incompressible stars are expected and have been
observed (see, e.g., Fig.~9 of \citep{Yagi_14_PRD}). However,
there is not much uncertainty in the EOS of low-density NS nuclear
matter, which is well studied. Such deviations are unimportant and
merely lead to modifications of the universal formulas instead of
breaking them. On the other hand, NSs and QSs behave differently
in the low-mass limit and hence  do not follow the same universal
formulas. Researchers could make use of the difference in the
universal trends obeyed, respectively, by NSs and QSs in the
low-mass limit to distinguish these two kinds of compact stars.

Two kinds of universal relations have been studied here, namely
the diagonal and the off-diagonal ones. In the former case,  the
scaled $f$-mode frequency $\bar{\omega}_l$ and the scaled tidal
deformability $\bar{\lambda}_l$ with the same angular momentum
index $l$ are linked together in an almost EOS-independent way. In
the latter case, $\bar{\omega}_l$ is related to $\bar{\lambda}_l'$
with $l \ne l'$. However, as shown in
Figs.~\ref{fig:k2_realistic}-\ref{fig:k5_realistic}, such
off-diagonal relations usually display stronger EOS dependence. On
the other hand, by combining these diagonal and off-diagonal
relations, other off-diagonal relations, such as $\bar{\omega}_l$
against $\bar{\omega}_{l'}$, or $\bar{\lambda}_l$ against
$\bar{\lambda}_{l'}$, with $l \ne l'$ can be readily established.

To put our work into proper perspective, we note that several
recent studies have been performed to extend the $I$-Love-$Q$
universality to multipolar sectors. For example,
Yagi~\cite{Yagi_14_PRD} found that there exists certain degree of
correlation between two tidal deformabilities with different
angular momentum indices albeit with more obvious EOS dependence.
On the other hand, in an attempt to generalize the no-hair theorem
for black holes to NSs (or QSs), the so-called ``three-hair
theorem" has been proposed
\citep{Yagi_hair_GR,Stein_hair_apj,practical-no-hair}, which
states that  higher multipole moments can be expressed in terms of
just the mass monopole, spin current dipole, and mass quadrupole
moments through EOS-independent relations. However, the accuracy
of the three-hair theorem was also found to deteriorate with the
order of multipole. Using the terminology coined here, we note
that these two relations are actually off-diagonal ones, where
more significant dependence on EOS is expected according to our
analysis.

\begin{figure}[h!]
  \centering
    \includegraphics[scale=0.3]{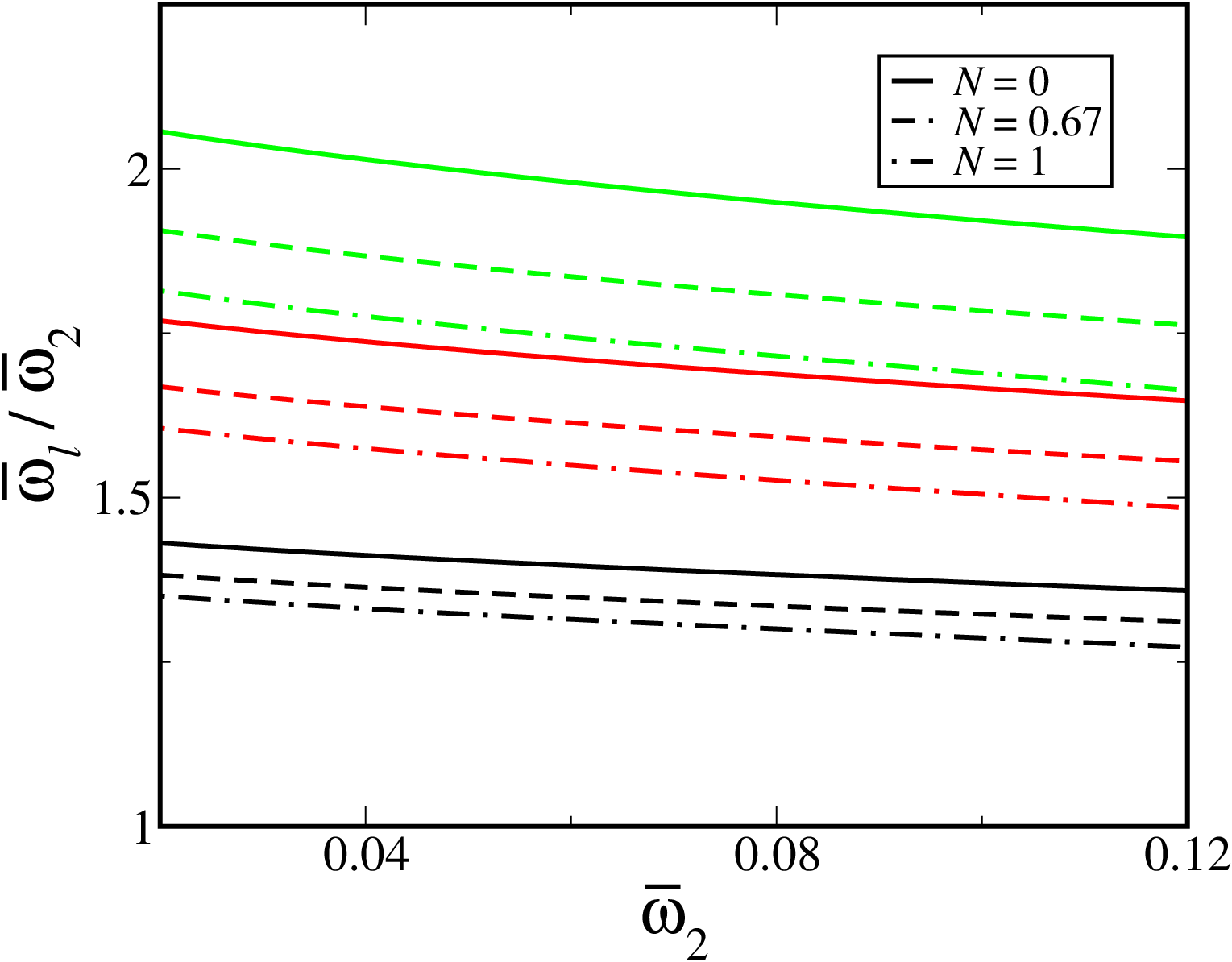}
   \caption{Plot of $\bar{\omega}_l/\bar{\omega}_2$ against $\bar{\omega}_2$ with  $l = 3$
   (black lines),  $l = 4$ (red lines), and $l
= 5$ (green lines) for incompressible
   stars with $N=0$ (solid lines),
   and polytropic
stars  with $N =0.67$ (dashed lines) and $N=1.0$ (dot-dashed
lines). }
   \label{fig:ff}
\end{figure}

For comparison,  we  show here  the other off-diagonal relation
which connects $\bar{\omega}_l$ of two different $l$ in
Fig.~\ref{fig:ff}, where $\bar{\omega}_l/\bar{\omega}_2$
($l=3,4,5$) is plotted against $\bar{\omega}_2$ for incompressible
stars ($N=0$), and polytropic stars with $N =0.67, 1.0$. As
discussed above, EOS dependence of these off-diagonal relations
become more obvious, especially for cases with larger difference
in the two angular momentum indices.

Finally, we remark that the present work is able to relate the
multipole moments considered in the three-hair theorem
 \citep{Yagi_hair_GR,Stein_hair_apj,practical-no-hair} to
 respective {\it f}-mode oscillation frequencies. Since the multipole
 moments of compact stars could be inferred by measuring the atomic spectra observed from  such stars with future x-ray telescopes such
 as LOFT and NICER \citep{LOFT,NICER}, the relevant {\it f}-mode oscillation frequencies
 could likewise be derived from the $f$-$I$ universal formula.


\section*{Acknowledgments}
We thank H.K. Lau and P.O. Chan for helpful discussions and the ideas
developed in their master theses.


\newcommand{\noopsort}[1]{} \newcommand{\printfirst}[2]{#1}
  \newcommand{\singleletter}[1]{#1} \newcommand{\switchargs}[2]{#2#1}

\end{document}